\documentclass[final,5p,twcolumns,sort&compress]{elsarticle}
\usepackage{amsmath}
\usepackage{widetext}
\usepackage[caption=false]{subfig}
\usepackage{graphicx}
\usepackage{epsfig}
\usepackage{amssymb}
\usepackage{multirow}
\usepackage{bigdelim}
\usepackage{slashed}
\usepackage{enumerate}
\usepackage{epstopdf}
\usepackage{epsfig}
\usepackage{enumerate}
\usepackage[colorlinks]{hyperref}
\usepackage{comment}
\usepackage[usenames, dvipsnames]{color}
\usepackage{hyperref}

\newcommand{\bea}{\begin{eqnarray}}
\newcommand{\eea}{\end{eqnarray}}
\newcommand{\bean}{\begin{eqnarray*}}
\newcommand{\eean}{\end{eqnarray*}}
\newcommand{\nn}{\nonumber \\}

\def\f{\tilde{f}}

\def\Feynarts{{{\sc FeynArts}}}
\def\Feyncalc{{{\sc FeynCalc}}}
\def\mathematica{{{\sc mathematica}}}
\def\sam{{{\sc S@M}}}

\newcommand{\QQ}{Q}
\newcommand{\GG}{G}

\newcommand{\GA}{\Gamma}

\def\Label#1{\label{#1}%
  \smash{\hbox to0pt{\raise1ex\hbox{\tiny[#1]}\hss}}}

\allowdisplaybreaks

\newcommand{\unipd}{Dipartimento di Fisica ed Astronomia, Universit\`a di Padova, Via Marzolo 8, 35131 Padova, Italy}
\newcommand{\pdinfn}{INFN, Sezione di Padova, Via Marzolo 8, 35131 Padova, Italy}
\newcommand{\mpi}{Max-Planck-Institut f\"ur Physik, F\"ohringer Ring 6, 80805 M\"unchen, Germany}
                      
\journal{Physics Letters B}
\begin{document}
\begin{frontmatter}
\title{Off-shell Currents and Color-Kinematics Duality}
\address[pd]{\unipd}
\address[infn]{\pdinfn}
\address[mpi]{\mpi}

\author[pd,infn,mpi]{Pierpaolo Mastrolia}
\ead{pierpaolo.mastrolia@cern.ch}
\author[pd,infn]{Amedeo Primo}
\ead{amedeo.primo@pd.infn.it}
\author[mpi]{Ulrich Schubert}
\ead{schubert@mpp.mpg.de}
\author[pd,infn]{William J. Torres Bobadilla}
\ead{william.torres@pd.infn.it}

\date{\today}

\begin{abstract}
We elaborate on the color-kinematics duality for off-shell diagrams in gauge theories coupled to matter, by investigating the scattering process $gg\to ss, q\bar q, gg$, and show that the Jacobi relations for the kinematic numerators of
off-shell diagrams, 
built with Feynman rules in axial gauge, reduce to a color-kinematics
violating term due to the contributions of sub-graphs only.
Such anomaly vanishes when the four particles connected by the
Jacobi relation are on their mass shell with vanishing squared momenta,
being either external or cut particles, 
where the validity of the color-kinematics duality is recovered. We
discuss the role of the off-shell decomposition in the direct
construction of higher-multiplicity numerators satisfying
color-kinematics identity in four as well as in $d$ dimensions, for
the latter employing the Four Dimensional Formalism variant of the Four Dimensional
Helicity scheme. We provide explicit examples for the QCD process $gg\to q\bar{q}g$.
\end{abstract}
\end{frontmatter}
\input{feynarts.sty}

\section{Introduction}

\begin{figure}[h]
\includegraphics[scale=1.05]{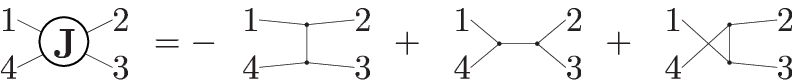}
\caption{The Jacobi combination: it can be applied both to Lie group
  structure constants and to the kinematic part of numerators.}\label{Fig:Jacobi:tree}
\end{figure}

Tree-level amplitudes in gauge theories are found to admit 
a {\it color-kinematics} (C/K) dual representation in terms of
diagrams involving only cubic vertices, 
where the kinematic part of the numerators obey Jacobi identities
and anti-symmetry relations similar to the ones holding for the
corresponding structure constants of the Lie algebra~\cite{Bern:2008qj,Bern:2010ue}, 
as depicted in Fig.\ref{Fig:Jacobi:tree}.

While first studies dealt with the C/K duality within scattering amplitudes involving massless partons,
more recent investigations pointed to the possibility that such a
symmetry can be present also when massive particles are involved~\cite{Johansson:2014zca,Naculich:2014naa,Johansson:2015oia}.

The C/K duality, on the one hand, implies the existence of 
relations between color ordered tree amplitudes in gauge theories
and, on the other hand,
yields a {\it gauge-gravity} dual representation of gravity
amplitudes, according to which they can be expressed as Yang-Mills amplitudes 
where the gauge-group structure constants are replaced by a second copy of the color-kinematics dual numerator,~\cite{Bern:2011ia,Bern:2010yg,Bern:2013yya,
BjerrumBohr:2009rd,BjerrumBohr:2010zs,BjerrumBohr:2010yc,BjerrumBohr:2010hn, BjerrumBohr:2012mg,
Stieberger:2009hq,Mafra:2010jq,Mafra:2011kj,Mafra:2011nv,
Grassi:2011ky,
Sondergaard:2011iv,
Feng:2010my,Jia:2010nz,Chen:2011jxa,Du:2011js,Du:2011se,Fu:2012uy,Du:2013sha,
Vaman:2010ez,
Boels:2011mn,Boels:2012ew,Boels:2013bi,
Oxburgh:2012zr,Saotome:2012vy,Broedel:2011pd,Broedel:2012rc,Cachazo:2012uq}.
When considering multi-loop amplitudes, the C/K duality
allows to establish relations between the (numerators of the)
integrands of planar and non-planar diagrams. Therefore, it 
turns into an efficient algorithm for generating either
high-multiplicity tree-level amplitudes or multi-loop
integrands, with a better control of the factorial growth of the
diagram complexity.

The C/K dual representation was proven
at tree-level by employing both string theory methods~\cite{BjerrumBohr:2009rd,BjerrumBohr:2010zs,Stieberger:2009hq,Mafra:2010jq,Mafra:2011kj,Mafra:2011nv} and on-shell methods~\cite{Feng:2010my,Jia:2010nz,Chen:2011jxa},
and it was conjectured to hold at higher orders~\cite{Bern:2010ue}.

Finding a systematic algorithm to determine C/K-dual 
numerators is not an easy task, because of the wide range of
transformations, referred to as \textit{generalized gauge
  invariance}, underpinning their
representation~\cite{Bern:2010yg,Monteiro:2011pc,BjerrumBohr:2012mg,Fu:2012uy,Du:2013sha}. 
In fact, the search for a C/K-dual representations can be formulated algebraically, at
least for tree-level numerators, in terms of an inverse linear
problem, as recently discussed in~\cite{Boels:2012sy,Carrasco:2015iwa}.
Nevertheless, it was possible to build an effective Lagrangian
with the property of generating C/K dual numerators for
tree-amplitudes~\cite{Tolotti:2013caa}, and non-trivial
examples of dual representation of higher-order numerators were found, up to two loops in non-supersymmetric theories,~\cite{Du:2012mt,Bern:2013yya,Nohle:2013bfa}, and up to four loops in supersymmetric ones,~\cite{Bern:2010ue,Bern:2012uf,Carrasco:2011mn,Bjerrum-Bohr:2013iza,Carrasco:2012ca,Bern:2011rj,BoucherVeronneau:2011qv,Mastrolia:2012wf,Schubert:2014paa}. \\ 

\begin{figure}[htb!]
\includegraphics[scale=1]{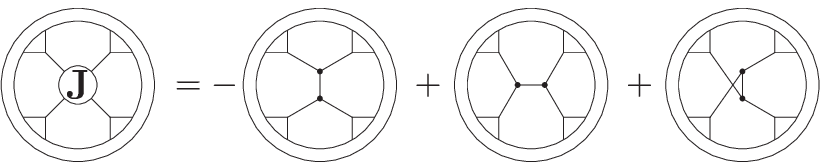}
\caption{Embedding of the Jacobi combination into either 
  higher-point or multi-loop diagrams.}\label{Fig:Jacobi:loop}
\end{figure}
 
In this letter, we study the role of color-kinematics duality 
within off-shell currents, which enter the construction of both
higher-multiplicity tree-level and multi-loop amplitudes.  
We investigate, in a purely diagrammatic approach, the origin of
possible deviations from the C/K-dual behavior, providing concrete
evidence of their relation to contact interactions, which was already
pointed out to in~\cite{Bern:2008qj,Tye:2010dd}.

First, we consider the tree-level diagrams for $gg \to X$, for
massless final state particles, with 
$X=s s, q {\bar q}, g g $, in four dimensions. 
We work in axial gauge, describing
scalars in the adjoint representation, 
while fermions in the fundamental one. 
We deal with the Jacobi relation of the kinematic numerators keeping
the partons off-shell.
Due to the off-shellness of the external particles, the C/K-duality is broken, and 
an {\it anomalous} term emerges. 
This anomaly vanishes in the on-shell massless limit, as it should, recovering the exact C/K-duality.

\par Later, we show that when the Jacobi combination of numerators is
immersed into a richer topology, associated either to higher-point
tree graphs or to loop integrands, as depicted in Fig.~\ref{Fig:Jacobi:loop},
the anomaly corresponds to the contribution of subdiagrams, obtained by pinching the external lines of the Jacobi combination.
In other words,
the Jacobi relation for the numerators in axial gauge, 
which, in the case of on-shell tree-amplitudes, is identically zero
and resolves the C/K-duality, 
in the off-shell case, can be expressed in terms of contact
interactions, 
which we explicitly identify for the first time for the
processes at hand and
represent one of the main result of this communication.
This decomposition, developed in the canonical formalism of Feynman diagrams, 
shows that the C/K-duality for high-multiplicity diagrams naturally holds when the
four particles entering the Jacobi combination are cut, 
since, in this case, the contribution of the subdiagrams trivially vanishes.

\par We discuss how our result, which provides a precise
identification of the anomalies which should be absorbed into the
redefinition of the trivalent numerators, can be used, together with
generalized gauge transformations ~\cite{Tye:2010dd,Bern:2008qj}, in
order to re-shuffle contact terms between diagrams and build on-shell
C/K-dual representations for higher-point tree-level amplitudes. As
successful check of this recursive construction, we present the
explicit calculation for the tree-level contribution to $gg\to
q\bar{q}g$.

Finally, we extend the C/K-duality to dimensionally regulated
tree-level amplitudes. They are the basic building
blocks for the determination of scattering amplitudes beyond
tree-level within generalized unitarity based methods, which require trees depending on the regulating parameter.

We adopt a novel variant of the four-dimensional helicity
(FDH) scheme~\cite{Bern:1991aq,Bern:1995db,Bern:2002zk}, the so-called
four dimensional formulation (FDF), recently proposed by some of the
authors in~\cite{Fazio:2014xea}. FDF has the advantages of employing
a purely four-dimensional representation of the additional degrees of
freedom which naturally enters when the space-time dimensions are
continued beyond four.
We derive the C/K relation for the basic four-point currents, and
determine the dual numerators for the process $gg\to
q\bar{q}g$ where the initial gluons live in $d$ dimensions.
\\ 

Algebraic manipulations and numerical evaluations have been
carried out by using the 
{\mathematica} packages \Feyncalc~\cite{Mertig:1990an} and
\sam~\cite{Maitre:2007jq}.


\section{Color-kinematics duality for scalars}
\label{sec:scalars}
The process $gg \to ss$ gets contributions from four tree-level diagrams, 
three of which contain cubic interactions, due to either $ggg$ or $gss$ couplings, while one is given by the quartic vertex $ggss$. 
Their color factors, for which we adopt the normalization $\f^{abc}=\text{Tr}([T^a,T^b]T^c)=i\sqrt{2}f^{abc}$, satisfy the Jacobi identity
\begin{align}
-\f^{a_1a_2b}\f^{ba_3a_4} +\f^{a_4a_1b}\f^{ba_2a_3} +\f^{a_4a_2b}\f^{ba_3a_1}=0 \ .
\label{jacobiG}
\end{align}
\noindent
A similar relation can be established for the kinematic part of the numerators of suitably defined graphs involving only cubic vertices. 

\begin{figure}[t]
\vspace*{0.5cm}
\includegraphics[scale=1.15]{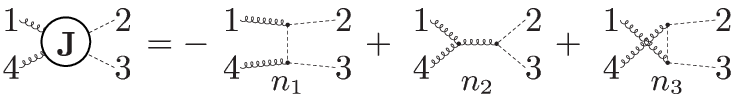}
\caption{Jacobi combination for $gg \to ss$.}
\label{BCJ1}
\end{figure}

In fact, after performing the color decomposition and some algebraic manipulations, 
the contribution of the four-point vertex can be distributed to the numerators of the 
diagrams with cubic vertices only, hence 
yielding the identification of three color-kinematics dual diagrams~\cite{Bern:2008qj}. 
The corresponding numerators, say $n_1$, $n_2$ and $n_3$, can be combined in Jacobi-like fashion, 
\begin{align}
N_{\text{s}}=-n_1+n_2+n_3, \label{BCJ}
\end{align}
as shown in Fig.\ref{BCJ1}.

In axial gauge - that we will consider throughout our calculations - the numerator of the gluon propagator takes the form
\begin{equation}
\Pi^{\mu\nu}(p,q)=\Pi^{\mu\nu}_{\text{Fey}}+\Pi^{\mu\nu}_{\text{Ax}}(p,q),
\end{equation}
where $\Pi^{\mu\nu}_{\text{Fey}}$ corresponds to the numerator of the propagator in Feynman gauge and $\Pi^{\mu\nu}_{\text{Ax}}(p,q)$ labels the term depending on an arbitrary light-like reference momentum $q^{\mu}$,
\begin{align}
\Pi^{\mu\nu}_{\text{Fey}}&=-i\,g^{\mu\nu}\,,\\
\Pi^{\mu\nu}_{\text{Ax}}(p,q)&=i\,\frac{p^\mu q^\nu+q^\mu p^\nu}{q\cdot p}.
\end{align}
The explicit form of (\ref{BCJ}) is given by the contraction of an
off-shell current with gluon polarizations as\footnote{From now on,
  the coupling constant as well as any constant prefactors associated to the normalization of the generators of the gauge group are understood.},
\begin{align}
N_{\text{s}}&=\big(J^{\mu_{1}\mu_{4}}_{\text{s-Fey}}+J_{\text{s-Ax}}^{\mu_{1}\mu_{4}}\big)\varepsilon_{\mu_{1}}\left(p_{1}\right)\varepsilon_{\mu_{4}}\left(p_{4}\right)\nn &=N_{\text{s-Fey}}+N_{\text{s-Ax}},
\label{BCJtg}
\end{align}
where $J_{\text{s-Fey}}^{\mu_{1}\mu_{4}}$ is the sum of the Feynman gauge-like terms of the three numerators,
\begin{align}
-i\,J_{\text{s-Fey}}^{\mu_{1}\mu_{4}}(p_1,p_2,p_3,p_4)=&p_1^{\mu _1} \left(p_1^{\mu _4}+2 p_3^{\mu _4}\right)\nn
&-\left(2 p_3^{\mu _1}+p_4^{\mu _1}\right) p_4^{\mu _4}
\label{jsfey}
\end{align}
and
\begin{multline}
-i\, J_{\text{s-Ax}}^{\mu_{1}\mu_{4}}(p_{1},p_{2},p_{3},p_{4})=\frac{1}{q\cdot(p_{1}+p_{4})}\bigg\{\\-\left(p_{1}^{\mu_{1}}p_{1}^{\mu_{4}}-p_{4}^{\mu_{1}}p_{4}^{\mu_{4}}-\left(p_{1}^{2}-p_{4}^{2}\right)g^{\mu_{1}\mu_{4}}\right)q\cdot\left(p_{2}-p_{3}\right)\\+\left(p_{2}^{2}-p_{3}^{2}\right)[\left(p_{4}+2p_{1}\right)^{\mu_{4}}q^{\mu_{1}}+q\cdot\left(p_{4}-p_{1}\right)g^{\mu_{1}\mu_{4}}\\-\left(p_{1}+2p_{4}\right)^{\mu_{1}}q^{\mu_{4}}]\bigg\}
\label{jsax}
\end{multline}
is the contribution, depending on the reference momentum, which only originates from $n_2$.\\

These expressions, obtained using only momentum conservation $\sum_{i=1}^{4}p_i=0$, show that the Jacobi identity holds also on the kinematic side, \textit{i.e.} $N_{s}=0$, once we impose the on-shell conditions of the four external particles, $p_{i}^2=0$, as well as the transversality condition for gluons, $p_i\cdot\varepsilon(p_i)=0\textrm{,}\quad i=1,4$.\\
We want to remark that $N_{\text{s-Fey}}$ and $N_{\text{s-Ax}}$ vanish separately, so that the C/K duality is satisfied at tree-level also in ordinary Feynman gauge.\\
A similar calculation was performed in~\cite{Zhu:1980sz}, where tree-level numerators for $gg\to X$ were studied as well. Our result differs in the choice of axial gauge, which, as we are going to show, plays an important role in the identification of the C/K-duality violating terms in the numerator higher-multiplicity graphs. \\

The expressions of the currents in Eqs.~(\ref{jsfey},\ref{jsax}) are valid for off-shell kinematics. Therefore,
they can be exploited for providing a better understanding of C/K-duality within more complex numerators obtained by embedding the Jacobi-like combination of tree-level numerators into a generic diagram, as depicted in Fig.~\ref{Fig:Jacobi:loop}, where the double circle shall represent an arbitrary number of loops and external legs.

\par\noindent In the most general case, the legs $p_1,p_2,p_3$ and $p_4$ become internal lines and polarizations associated to the particles are replaced by the numerator of their propagators, which, for the scalar case, simply corresponds to a factor $i$.
Accordingly, Eq.~\eqref{BCJ} generalizes to the following contraction,
\begin{align}
N_{\text{s}} &=(N_{\text{s}})_{\alpha_1\alpha_4}X^{\alpha_1\alpha_4}.
\label{eq:Ns:def}
\end{align}
between the tensor $(N_{\text{s}})_{\alpha_1\alpha_4}$, defined as,
\begin{equation}
(N_{\text{s}})_{\alpha_1\alpha_4}=-
\big(J^{\mu_{1}\mu_{4}}_{\text{s-Fey}}+J_{\text{s-Ax}}^{\mu_{1}\mu_{4}}\big)
\Pi_{\mu_1\alpha_1}\!\!\left(p_1,q_1\right)
\Pi_{\mu_4\alpha_4}\!\!\left(p_4,q_4\right).
\label{BCJgN}
\end{equation}
and the arbitrary tensor $X^{\alpha_1\alpha_4}$, standing for the residual kinematic dependence of the diagrams, associated to either higher-point tree-level or to multi-loop topologies. 

\begin{figure}[t]
\vspace{0.5cm}
\includegraphics[scale=0.95]{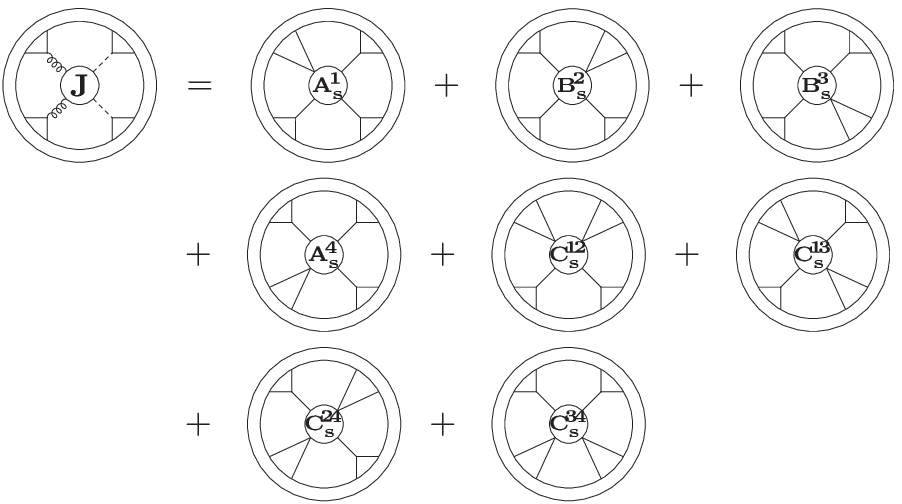}
\caption{Off-shell color-kinematics duality for gluons and scalars. The Jacobi combination of tree-level numerators (l.h.s)  is expressed in terms of subdiagrams only (r.h.s.). }
\label{bcjLoopa}
\end{figure}

Using momentum conservation, we find that the r.h.s. of \eqref{BCJgN} can be cast in the following suggestive form,
\begin{align}
(N_{\text{s}})_{\alpha_{1}\alpha_{4}}=&\,\, p_{1}^{2}(A_{s}^{1})_{\alpha_{1}\alpha_{4}}+p_{4}^{2}(A_{s}^{4})_{\alpha_{1}\alpha_{4}}\nn
&+p_{2}^{2}(B_{s}^{2})_{\alpha_{1}\alpha_{4}}+p_{3}^{2}(B_{s}^{3})_{\alpha_{1}\alpha_{4}}\nn
&+p_{1}^{2}p_{2}^{2}(C_{s}^{12})_{\alpha_{1}\alpha_{4}}+p_{1}^{2}p_{3}^{2}(C_{s}^{13})_{\alpha_{1}\alpha_{4}}\nn
&+p_{4}^{2}p_{2}^{2}(C_{s}^{24})_{\alpha_{1}\alpha_{4}}+p_{4}^{2}p_{3}^{2}(C_{s}^{34})_{\alpha_{1}\alpha_{4}}\,,
\label{BCJgN1}
\end{align}
where $A^{i}_s, B^{i}_s$ and $C^{ij}_s$ are tensors depending both on the momenta $p_i$ of gluons and scalars, eventually depending on the loop variables, and on the reference momenta $q_i$ of each gluon propagators.

Remarkably, Eq.(\ref{BCJgN1}) shows the full decomposition of a generic numerator built from the Jacobi relation in terms of squared momenta of the particles entering the Jacobi combination defined in Fig.~\ref{BCJ1}. In particular,  this result implies that the C/K duality is certainly satisfied when imposing the on-shell cut-conditions $p_i^2 =0$. 

A diagrammatic representation of the consequences of the decomposition (\ref{BCJgN1}) in (\ref{eq:Ns:def}) is given in Fig.~\ref{bcjLoopa}, where the eight terms appearing in r.h.s. of (\ref{BCJgN1}) generate subdiagrams, obtained by pinching one or two denominators. In these subdiagrams $A^{i}_s$, $B^{i}_s$ and $C^{ij}_s$ play the role of effective vertices contracted with the tensor $X^{\alpha_1\alpha_4}$.\\

The existence of contact terms responsible for the violation of the C/K-duality was conceptually pointed out already in~\cite{Bern:2008qj}. Here, we identified, for the first time to our knowledge, on a purely diagrammatic basis, 
the sources of such \textit{anomalous} term, exposed in the (single and double) momentum-square dependance of 
formula (\ref{BCJgN1}). \\ 
The choice of axial gauge turned out to be crucial within our derivation, since the $p^2$-terms appear, beside from the trivial contraction $p^\mu p_{\mu}$, also from 
the contraction of $\Pi^{\mu\nu}(p,q)$ with the corresponding gluon momentum (Ward identity),
\begin{equation}
p_{\mu}\Pi^{\mu\nu}(p,q)=i\,p_\mu \bigg(\!\!-g^{\mu\nu}+\frac{p^\mu q^\nu+q^\mu p^\nu}{q\cdot p}\bigg)=i\,p^2\frac{q^\nu}{q\cdot p}.
\label{can}
\end{equation}

For the sake of simplicity, we do not provide the explicit expressions for $A^{i}_s$, $B^{i}_s$, and $C^{ij}_s$. \\
By inspection of \eqref{jsfey} and \eqref{jsax}, we observe that $J_{\text{s-Fey}}$ only gives contribution to $A^{1}_\text{s}$ and $A^{4}_\text{s}$, while $J_{\text{s-Ax}}$ produces terms proportional to the momenta of all the four particles as well as to all the possible pairs of gluon-scalar denominators, \textit{i.e.} contributes to all the eight effective vertices.\\
In addition, because of the explicit symmetries of  $J_{\text{s-Fey}}$ and  $J_{\text{s-Ax}}$  under $1\leftrightarrow4$ and $2\leftrightarrow 3$, the two effective vertices associated to the pinch of one scalar propagator, namely $A_\text{s}^{1}$ and $A_\text{s}^4$, are related to each other by particle relabelling. The same happens for $B_\text{s}^2$ and $B_\text{s}^3$, which correspond to the pinch of one gluon propagator. For the same reason, there is only one independent $C_s$ function, corresponding to the pinch of two denominators, say $p_i^2p_j^2$, which are originated from terms proportional to $p_i^\mu p_j^2$ in Eq.\eqref{jsax}. \\ 

In the following Sections, we show that off-shell color-kinematics identities can be established, along the same lines, for the coupling of gluons to quarks as well as for pure gauge interactions.



\section{Color-kinematics duality for quarks} 
\label{sec:quarks}
\begin{figure}[h]
\vspace*{0.5cm}
\includegraphics[scale=1.15]{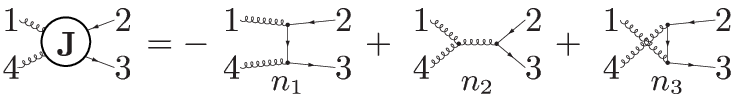}
\caption{Jacobi combination for $gg\to q\bar{q}$.}\label{BCJf}
\end{figure}

The tree-level scattering $gg\to q\bar{q}$ has a simpler diagrammatic
structure than the previously discussed case, because of the absence of
any four-particle coupling. There are three Feynman graphs
contributing to it, and they contain only cubic
interactions due to $ggg$ and $gq\bar{q}$ couplings.
The corresponding color factors obey the Jacobi identity,
\begin{equation}
-T^{a_4}_{3\bar{\j}}T^{a_1}_{\j\bar{2}}+T^{a_1}_{3\bar{\j}}T^{a_4}_{\j\bar{2}}+\f^{a_4a_1b}T^{b}_{3\bar{2}}=0,
\end{equation}
and, in this case, it is straightforward to build the combination of color-kinematics numerators for the tree-level
graph (\ref{BCJ}), as shown in Fig.~\ref{BCJf}. 

Following the same derivation as for gluons and scalars, the Jacobi
relation for gluons and fermions can be built from the contraction of
fermion currents and polarizations,
\begin{align}
N_{\text{q}}&=\bar{u}(p_3)({J}_\text{q-Fey}^{\mu_{1}\mu_{4}}+{J}_{\text{q-Ax}}^{\mu_{1}\mu_{4}}) v(p_2)\varepsilon_{\mu_4}(p_4)\varepsilon_{\mu_1}(p_1)\nn
&=N_{\text{q-Fey}}+N_{\text{q-Ax}}.
\label{num}
\end{align}

\noindent
Using Dirac algebra and momentum conservation,
$J_\text{q-Fey}^{\mu_{1}\mu_{4}}$ and
$J_{\text{q-Ax}}^{\mu_{1}\mu_{4}}$ can be organized into compact forms as,
\begin{align}
-i\,J_\text{q-Fey}^{\mu_{1}\mu_{4}}&(p_1,p_2,p_3,p_4)=\nn
&=-\slashed p_{3}\gamma^{\mu_{4}}\gamma^{\mu_{1}}-\gamma^{\mu_{1}}\gamma^{\mu_{4}}\slashed p_{2}+\left(\slashed p_{3}+\slashed p_{2}\right)g^{\mu_{1}\mu_{4}}\nn
&\quad+p_{4}^{\mu_{4}}\gamma^{\mu_{1}}-p_{1}^{\mu_{1}}\gamma^{\mu_{4}}
\label{BCJq}
\end{align}
and
\begin{multline}
-i\,J_{\text{q-Ax}}^{\mu_{1}\mu_{4}}(p_1,p_2,p_3,p_4)=\\
=\frac{1}{q\cdot(p_4+p_1)}\bigg\{ (\slashed{p}_3+\slashed{p}_2)\big[g^{\mu_1\mu_4}q\cdot(p_4-p_1)\\
-q^{\mu_4}(p_1+2p_4)^{\mu_1}+q^{\mu_1}(p_4+2p_1)^{\mu_4}\big]\\
-\slashed{q}\big[p_1^{\mu_4}p_1^{\mu_1}-p_4^{\mu_4}p_4^{\mu_1}+g^{\mu_1\mu_4}(p_4^{2}-p_1^{2})\big]\bigg\}
\label{BCJqax}.
\end{multline}

We observe that \eqref{num} vanishes when the four external particles
are on-shell, due to transversality conditions and Dirac equation, $\bar{u}(p_3)\slashed{p_3}=\slashed{p_2}v(p_2)=0$.\\
The C/K duality is satisfied at tree-level also in Feynman gauge, since on-shellness enforces $N_\text{q-Fey}$ and
$N_{\text{q-Ax}}$ to vanish independently.\\

In order to study the Jacobi combination within
higher-point numerators or multi-loop integrands, we
repeat the procedure adopted in Section~\ref{sec:scalars}. Accordingly, 
we promote the external states of gluons and quarks to propagating particles, and
define the off-shell tensor, 
\begin{align}
(N_{\text{q}})_{\alpha_1\alpha_4}&=
i\slashed p_{3}({J}_{\text{q-Fey}}^{\mu_{1}\mu_{4}}+{J}_{\text{q-Ax}}^{\mu_{1}\mu_{4}})i\slashed p_{2}\nn
&\quad\times\Pi_{\mu_{4}\alpha_{4}}\!\!\left(p_{4},q_{4}\right)
\Pi_{\mu_{1}\alpha_{1}}\!\!\left(p_{1},q_{1}\right),
\label{BCJqN}
\end{align}
by replacing the polarization vectors with the numerators of the gluon
propagators, 
and the spinors $\bar{u}(p_3)$ and $v(p_2)$ with the numerators of
fermionic propagators. 
As before, the full numerator is obtained contracting \eqref{BCJqN} with a generic tensor $X^{\alpha_1\alpha_4}$.\\

\begin{figure}[t]
\vspace{0.5cm}
\includegraphics[scale=0.95]{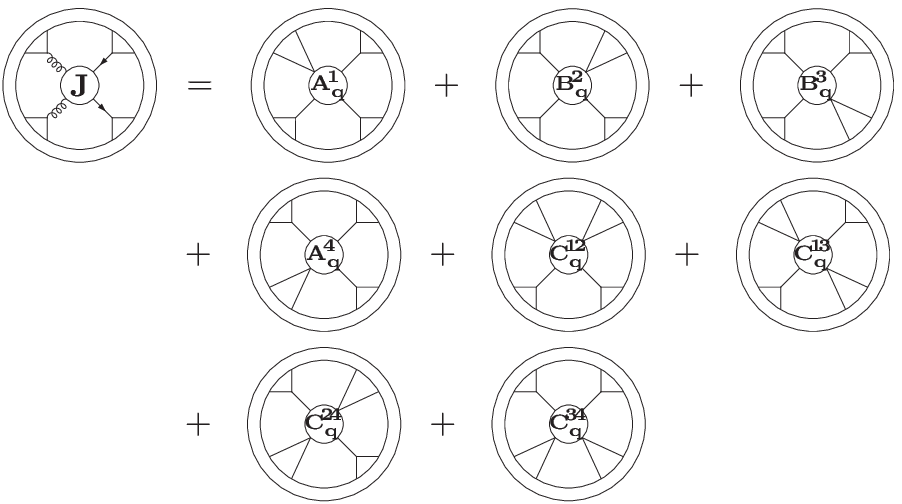}
\caption{Off-shell color-kinematics duality for gluons and quarks.}\label{bcjLoopb}
\end{figure}

Manipulating the r.h.s. of \eqref{BCJqN}, we obtain an expression
analogous to the decomposition (\ref{BCJgN1}), where the denominators
of the four particles are manifestly factored out,
\begin{align}
(N_{\text{q}})_{\alpha_{1}\alpha_{4}}=&p_{1}^{2}(A_{q}^{1})_{\alpha_{1}\alpha_{4}}+p_{4}^{2}(A_{q}^{4})_{\alpha_{1}\alpha_{4}}\nn
&+p_{2}^{2}(B_{q}^{2})_{\alpha_{1}\alpha_{4}}+p_{3}^{2}(B_{q}^{3})_{\alpha_{1}\alpha_{4}}\nn
&+p_{1}^{2}p_{2}^{2}(C_{q}^{12})_{\alpha_{1}\alpha_{4}}+p_{1}^{2}p_{3}^{2}(C_{q}^{13})_{\alpha_{1}\alpha_{4}}\nn
&+p_{4}^{2}p_{2}^{2}(C_{q}^{24})_{\alpha_{1}\alpha_{4}}+p_{4}^{2}p_{3}^{2}(C_{q}^{34})_{\alpha_{1}\alpha_{4}}\, .
\label{NqP}
\end{align}
In the above expression, $A^i_q$ and $B^i_q$ receive contribution both from $J_{\text{q-Fey}}$ and $J_{\text{q-Ax}}$ while $C^{ij}_q$'s are determined only by $J_{\text{q-Ax}}$.\\
This can be understood by inspection of \eqref{BCJq} and
\eqref{BCJqax}, observing 
that denominators may appear because of \eqref{can}, as well as
because of the identity $\slashed{p}\slashed{p}=p^2$. \\
Also in this case, we only have three independent functions: two for the effective vertices corresponding to the pinch of one quark- or one gluon- propagator, namely $A^i_q$ and $B^i_q$, and a single vertex $C^{ij}_q$ for the pinch of a quark-gluon pair. We remark that these functions contain non-trivial Dirac structure.\\
The interpretation of \eqref{NqP} is similar to the one of \eqref{BCJgN1} and it is illustrated in Fig.~\ref{bcjLoopb}.



\section{Color-kinematics duality for gluons}
\label{sec:gluons}
\begin{figure}[h]
\vspace*{0.5cm}
\includegraphics[scale=1.15]{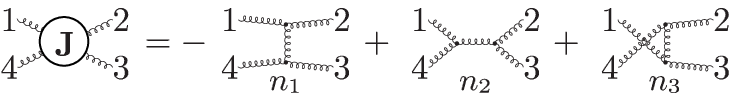}
\caption{Jacobi combination for gluons.}\label{BCJs}
\end{figure}

Finally, we consider the C/K duality for the pure gauge interaction process $gg \to gg$.\\
As for $gg\to ss$, there are four diagrams to be considered, three involving the tri-gluon interaction, and one 
containing the four-gluon vertex.
The color factors obey the Jacobi identity \eqref{jacobiG}.
After distributing the contribution of the four-gluon vertex into the three structures according to the color decomposition, we can define three graphs with cubic vertices only whose numerators enter 
a Jacobi combination, as shown in Fig.~\ref{BCJs}. 

With this prescription, the kinematic Jacobi identity takes the form
\begin{align}
N_{\text{g}}&=\big({J}_{\text{g-Fey}}^{\mu_{1}...\mu_{4}}+{J}_{\text{g-Ax}}^{\mu_{1}...\mu_{4}}\big)  \varepsilon_{\mu_1}(p_1) \varepsilon_{\mu_2}(p_2) \varepsilon_{\mu_3}(p_3)\varepsilon_{\mu_4}(p_4)\nn
&=N_{\text{g-Fey}}+N_{{\text{g-Ax}}},
\label{numS}
\end{align}
where  
\begin{multline}
-i\,{J}_{\text{g-Fey}}^{\mu_{1}\mu_{2}\mu_{3}\mu_{4}}(p_1,p_2,p_3,p_4)=\\
=p_{1}^{\mu_{1}}[g^{\mu_{3}\mu_{4}}\left(p_{1}+2p_{4}\right)^{\mu_{2}}-g^{\mu_{2}\mu_{4}}\left(p_{1}+2p_{4}\right)^{\mu_{3}}\\+g^{\mu_{2}\mu_{3}}\left(p_{1}+2p_{3}\right)^{\mu_{4}}]\\-p_{2}^{\mu_{2}}[g^{\mu_{3}\mu_{4}}\left(p_{2}+2p_{4}\right)^{\mu_{1}}-g^{\mu_{1}\mu_{4}}\left(p_{2}+2p_{4}\right)^{\mu_{3}}\\+g^{\mu_{1}\mu_{3}}\left(p_{2}+2p_{3}\right)^{\mu_{4}}]\\+p_{3}^{\mu_{3}}[g^{\mu_{2}\mu_{4}}\left(p_{3}+2p_{4}\right)^{\mu_{1}}-g^{\mu_{1}\mu_{4}}\left(p_{3}+2p_{4}\right)^{\mu_{2}}\\+g^{\mu_{1}\mu_{2}}\left(p_{3}+2p_{2}\right)^{\mu_{4}}]\\-p_{4}^{\mu_{4}}[g^{\mu_{2}\mu_{3}}\left(p_{4}+2p_{3}\right)^{\mu_{1}}-g^{\mu_{1}\mu_{3}}\left(p_{4}+2p_{3}\right)^{\mu_{2}}\\+g^{\mu_{1}\mu_{2}}\left(p_{4}+2p_{2}\right)^{\mu_{3}}]
\label{jfeyg}
\end{multline}
and
\begin{multline}
-i\, J_{\text{g-Ax}}^{\mu_{1}\mu_{2}\mu_{3}\mu_{4}}(p_{1},p_{2},p_{3},p_{4})=\frac{1}{q\cdot(p_{1}+p_{2})}\Bigg\{\\\left(p_{1}^{\mu_{1}}p_{1}^{\mu_{2}}-p_{2}^{\mu_{2}}p_{2}^{\mu_{1}}-\left(p_{1}^{2}-p_{2}^{2}\right)g^{\mu_{1}\mu_{2}}\right)[q\cdot\left(p_{4}-p_{3}\right)g^{\mu_{3}\mu_{4}}\\-\left(p_{3}+2p_{4}\right)^{\mu_{3}}q^{\mu_{4}}+\left(p_{4}+2p_{3}\right)^{\mu_{4}}q^{\mu_{3}}]\\+\left(p_{3}^{\mu_{3}}p_{3}^{\mu_{4}}-p_{4}^{\mu_{3}}p_{4}^{\mu_{4}}-\left(p_{3}^{2}-p_{4}^{2}\right)g^{\mu_{3}\mu_{4}}\right)[q\cdot\left(p_{1}-p_{2}\right)g^{\mu_{1}\mu_{2}}\\+\left(p_{1}+2p_{2}\right)^{\mu_{1}}q^{\mu_{2}}-\left(p_{2}+2p_{1}\right)^{\mu_{2}}q^{\mu_{1}}]\Bigg\}
\\-[(1234)\to(4123)]-[(1234)\to(4231)].\label{jaxg}
\end{multline}
With the by-now usual arguments, we observe that the tree-level C/K duality, $N_{\text{g-Fey}}=N_{{\text{g-Ax}}}=0$, holds when the external particles are on-shell, separately for the Feynman- and axial-gauge contributions.\\

\begin{figure}[h]
\includegraphics[scale=0.96]{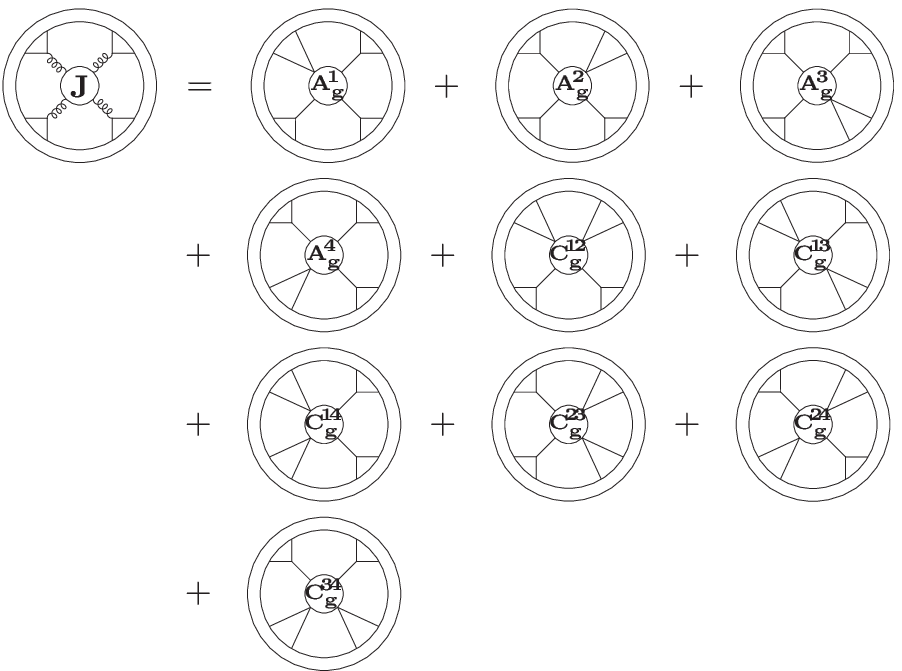}
\caption{Off-shell color-kinematics duality for gluons.}\label{bcjLoopc}
\end{figure}

As in the previous Sections, we can build a generic off-shell tensor, to be embedded in a more complex topology, either with more loops or more legs, by replacing polarization vectors with the numerator of the corresponding propagators, 
\begin{multline}
\left(N_{\text{g}}\right)_{\alpha_1...\alpha_4}\!=
\big(J^{\mu_{1}..\mu_{4}}_{\text{g-Fey}}+J^{\mu_{1}...\mu_{4}}_{\text{g-Ax}}\big) 
\Pi_{\mu_1\alpha_1}\!\!\left(p_1,q_1\right)\\ 
\Pi_{\mu_2\alpha_2}\!\!\left(p_2,q_2\right)
\Pi_{\mu_3\alpha_3}\!\!\left(p_3,q_3\right)\Pi_{\mu_4\alpha_4}\!\!\left(p_4,q_4\right),
\label{gluonnum}
\end{multline}
and by contracting this expression with an appropriate tensor $X^{\alpha_1...\alpha_4}$.\\
 Because of the Ward identity \eqref{can}, $\left(N_{\text{g}}\right)_{\alpha_1...\alpha_4}$ turns out to be decomposed as
\begin{align}
\left(N_{\text{g}}\right)_{\alpha_1...\alpha_4}\!&=
\sum_{i=1}^{4}p_i^2(A^i_g)_{\alpha_1...\alpha_4}+\sum_{\substack{i,j=1\\
		i\neq j}}^{4}p_i^2p_j^2(C^{ij}_g)_{\alpha_1...\alpha_4}.
\label{NsP}
\end{align}

\noindent
We observe that, differently from the scalar and fermionic cases, $J_{\text{g-Ax}}$ produces all the possible combinations of two different denominators. This is a consequence of the permutation symmetry of the gauge-dependent part of the numerators, which is an exclusive feature of the Jacobi identity for pure gauge interactions. The same symmetry reduces from three to two the number of independent effective vertices, corresponding to the pinch of one or two gluon propagators.\\
The diagrammatic effect of \eqref{NsP} contracted with $X^{\alpha_1...\alpha_4}$ is depicted in  Fig.~\ref{bcjLoopc}, 
which shows that, as it happened for the $gg \to ss$, and $gg \to q {\bar q}$,
the Jacobi combination of the kinematic numerators for $gg \to gg$ with off-shell particles always reduces to subdiagrams. 

Let us, finally, remark that the form factors $A^i$-, $B^i$- and $C^{ij}$-type appearing in the decompositions 
\eqref{BCJgN1}, \eqref{NqP}, and \eqref{NsP} still depend on the momenta $p_i$ and $p_j$. Therefore, within multiloop integrands, they can generate tensor integrals which can be subject to further integral reductions.



 \section{Construction of dual numerators for higher-point amplitudes}
\label{method}
In this Section we illustrate how the previous results can be used, together with \textit{generalized gauge invariance},~\cite{Bern:2010ue,Bern:2010yg,Tye:2010dd,Bern:2008qj}, in order to explicitly determine dual representation of higher-point amplitudes starting from Feynman diagrams. In addition, we show that our construction allows a purely diagrammatic derivation of monodromy relations for amplitudes, ~\cite{BjerrumBohr:2010zs}.\\
Any tree-level $m$-point amplitude can be decomposed in terms of cubic diagrams
\begin{align}
\mathcal{A}^{\text{tree}}_{m}(p_1,p_2, ..., p_m)=\sum_{i=1}^{N}\frac{c_in_i}{D_i},
\label{dressed}
\end{align}
where $c_i$ is the color factor associated to the $i^{th}$-graph, $D_i$ collects the denominators of all internal propagators and $n_{i}$ is the kinematic numerator which, besides the appropriate Feynman rule-term, might contain contributions from contact interactions, that are assigned with the prescription described in Sections~\ref{sec:scalars} and \ref{sec:gluons}.\\
\par The $N$ color factors appearing in \eqref{dressed}, satisfy a set of $M$, $M<N$, Jacobi identities
\begin{align}
-c_i+c_j+c_k=0,
\label{jaccol}
\end{align}
whose solution allows us to express $M$ color factors in terms of $N-M$ independent ones $\{c_{\sigma(1)},...,c_{\sigma(N-M)}\}$, so that \eqref{dressed} can be organized as\footnote{
	The amplitude decomposition in terms of a color basis
	obtained as a direct solution of the Jacobi/Lie algebra relations is
	equivalent, yet different from the decomposition proposed
	in~\cite{DelDuca:1999rs}. In particular, the kinematic coefficient $K_{\sigma(i)}$ of
	each color factor appearing in~\eqref{dec3} is gauge invariant, and
	corresponds, in general, to a linear combination of color ordered partial amplitudes.}
\begin{align}
\mathcal{A}^{\text{tree}}_{m}(p_1,p_2, ..., p_m)&=\sum_{i=1}^{N-M}c_{\sigma(i)}K_{\sigma(i)}(\{n\}),\label{dec3}
\end{align}
where
	\begin{align}
	K_{\sigma(i)}(\{n\})=\sum_{j=1}^{N}\frac{\alpha_{\sigma(i)j}n_j}{D_j}, \quad  \alpha_{\sigma(i)j}\in\{0,\pm1\}.
	\label{kinfactor}
	\end{align}
The set of identities \eqref{jaccol} is not, in general, trivially satisfied by the corresponding kinematic numerators, whose Jacobi combinations produce non-vanishing anomalous terms,
\begin{align}
-n_i+n_j+n_k=\phi_{[i,j,k]}.
\label{anomaly}
\end{align}
We observe  that the two sets of equations \eqref{kinfactor} and \eqref{anomaly} can be conveniently organized into the matrix equation,
\begin{align}
\mathbb{A}\,\mathbf{n}=\mathbf{K}+\boldsymbol\phi,
\label{systemn}
\end{align}
with 
\begin{align}
&\mathbf{n}=(n_1,n_2, ..., n_{N})^{T},\nn
&\mathbf{K}=(\{K_{\sigma(i)}\},0,0,...,0)^{T},\nn
&\boldsymbol\phi=(0,0,...,0,\{\phi_{[i,j,k]}\})^{T}
\label{vector}
\end{align}
and
\begin{align}
\mathbb{A}_{ij}&\in\{0,\pm1,\pm{D_j}^{-1}\}. 
\label{matA}
\end{align}
As shown by the decompositions \eqref{BCJgN1}, \eqref{NqP} and \eqref{NsP}, the anomalies are proportional to the off-shell momenta of the particle entering the Jacobi combination itself. Therefore, the rise of these anomalies seems to be related to the allocation of contact terms between cubic diagrams, which naturally provides numerators satisfying C/K-duality in the four-point case only.\\
As a consequence, in order to obtain a dual representation of the amplitude, we need to re-shuffle contact terms, leaving \eqref{dressed} unchanged. This can be achieved through a generalized gauge transformation, which consists in a set of shifts of the kinematic numerators,
\begin{align}
n_i\to n'_i=n_i-\Delta_i,
\label{shifts}
\end{align}
satisfying

\begin{align}
\delta\mathcal{A}^{\text{tree}}_{m}(p_1,p_2, ..., p_m)\equiv\sum_{i=1}^{N-M}c_{\sigma(i)}\sum_{j=1}^{N}\frac{\alpha_{\sigma(i)j}\Delta_j}{D_j}=0,
\label{gaugeinv}
\end{align}
in such a way that the amplitude can still be written as
\begin{align}
\mathcal{A}^{\text{tree}}_{m}(p_1,p_2, ..., p_m)=\sum_{i=1}^{N}\frac{c_in'_i}{D_i}=\sum_{i=1}^{N-M}c_{\sigma(i)}K_{\sigma(i)}(\{n'\}).
\label{newamp}
\end{align}
By imposing the vanishing of the coefficient of each $c_{\sigma(i)}$ in \eqref{gaugeinv}, the gauge invariance requirement is translated into a set equations for the shifts, which leaves $M$ of them undetermined. This means that, in principle, we have enough freedom to ask the shifts to be solution of $M$ additional equations,
\begin{align}
-\Delta_i+\Delta_j+\Delta_k=\phi_{[i,j,k]},
\label{jacshift}
\end{align}
which, inserted in \eqref{anomaly}, make the new set of numerators $n'_i$ manifestly dual.\\
Thus, the simultaneous imposition of \eqref{gaugeinv} and \eqref{jacshift} leads the determination of numerators satisfying the C/K-duality back to the solution of the $N \times N$ linear system
\begin{align}
\mathbb{A}\,\Delta=\boldsymbol\phi,
\label{system}
\end{align}  
with 
\begin{align}
&\Delta=(\Delta_1,\Delta_2, ..., \Delta_{N})^{T},
\end{align}
whereas the vector $\boldsymbol\phi$ and the matrix $\mathbb{A}$  are the ones defined by \eqref{vector} and \eqref{matA}.\\
\par By solving \eqref{system}, we can determine the shifts to be performed on the numerators obtained from Feynman diagrams ensuring C/K-duality as a function of anomalies $\phi_{[i,j,k]}$ and denominators $D_{i}$.\\ 
We note that the existence of a dual representation of the amplitude is bound to the consistency of the non-homogenous system \eqref{system}, \textit{i.e.} to the condition
\begin{align}
\text{rank}(\mathbb{A}|\boldsymbol\phi)=\text{rank}(\mathbb{A}),
\label{cond}
\end{align}
where $\mathbb{A}|\boldsymbol\phi$ is the augmented matrix associated to $\mathbb{A}$.\\
In particular, if the system had maximum rank $N$, the expression of the numerators would be completely fixed by C/K-duality.\\
However, as we will show in an explicit example, the rank of the system turns out to be smaller than $N$, so that its solution will 
depend on a set of arbitrary shifts, which are left completely undetermined by the imposition of C/K-duality. The existence of a residual freedom in the choice of the
dual representation was first observed in
~\cite{Bern:2008qj} and  more recently, in ~\cite{Boels:2012sy}, where the reduction of the tree-level C/K-duality to an underconstrained linear problem is addressed 
in terms of a \textit{pseudo-inverse} operation, it has 
been interpreted as the hint of a possible analogous construction at loop-level.\\
\\
We observe that, if
\begin{align}
\text{rank}(\mathbb{A})<N,
\label{under}
\end{align}
the condition \eqref{cond} can be satisfied only if a number $N-\text{rank}(\mathbb{A})$ of relations can be established between the anomalous terms $\phi_{[i,j,k]}$.
\\In the following Section, we will show that these constraints, which were obtained in \cite{BjerrumBohr:2010zs} for the five-gluon amplitude using string-derived monodromy relations, are fully implied by the diagrammatic expansion of the amplitude
and that they can be obtained from the simple knowledge of the matrix $\mathbb{A}$. In particular, they are found by determining a complete set of vanishing linear combinations of rows of $\mathbb{A}$. Working on a specific example we will argue that our off-shell decomposition \eqref{BCJgN1},\eqref{NqP} and \eqref{NsP} make these relations manifest.
\\Moreover, we observe that, because of \eqref{systemn} and \eqref{system}, applying the matrix $\mathbb{A}$ to the new set of numerators $n'_i$, we obtain
\begin{align}
\mathbb{A}\,\mathbf{n}'=\mathbf{K},
\label{systemnh}
\end{align}
with $\mathbf{K}$ given by \eqref{vector}.\\
Again,if $\text{rank}(\mathbb{A})<N$, the consistency condition
\begin{align}
\text{rank}(\mathbb{A}|\mathbf K)=\text{rank}(\mathbb{A}),
\label{condn}
 \end{align}
implies the existence of $N-\text{rank}(\mathbb{A})$ constraints between the kinematic factors $K_{\sigma(i)}$, which are in one-by-one correspondence with the relations between color ordered amplitudes, that have been previously conjectured as an implication of of C/K-duality  ~\cite{Bern:2008qj}, and then derived from the low energy limit of string theory ~\cite{BjerrumBohr:2010zs} and on-shell recursion~\cite{Feng:2010my}.\\
Therefore, this construction shows that all these non-trivial relations can be derived from the expansion of the amplitudes in terms of Feynman graphs through purely algebraic manipulation on the matrix $\mathbb{A}$.\\

Finally, we want to remark that, whereas in the usual top-down approach, the numerators appearing in the r.h.s. of \eqref{dressed} are interpreted as abstract re-organization of Feynman rules-numerators on which, by assumptions, the C/K-duality is imposed, our approach provides a systematic way to identify the link between dual numerators and Feynman diagrams. Starting from the set of explicitly C/K-violating but well-defined Feynman rule-numerators we determine, by mean of generalized gauge transformations, the actual redistribution of contact terms which has to be performed in order to establish the duality. \\
In this framework, the off-shell decomposition of the four-point identities derived in the previous sections plays a key role in the identification and in the algorithmic construction of the anomalous terms, \textit{i.e.} the non-vanishing element of the vector $\boldsymbol\phi$. In fact, the l.h.s. of each kinematic equation \eqref{anomaly} can be obtained from the contraction of $(N_{k})_{\alpha_{i}...\alpha_{j}}$ ($k=\text{s,q,g}$, depending on the process under consideration) evaluated on a suitable permutation of the labels of the external legs, with lower-point functions. Therefore, all the anomalies $\phi_{[i,j,k]}$ can be determined, without going trough the explicit calculation of all $N$ numerators, just by identifying the $M$ tree-level subdiagrams that can be factored in each of three numerators appearing in \eqref{anomaly}. \\
\\
Summarizing the diagrammatic approach to the construction of C/K-dual numerators for higher-point amplitudes:
\begin{itemize}[-]
\item given the decomposition of an amplitude in terms of Feynman diagrams, it is organized into $N$ cubic graphs, whose numerators satisfy the system of equations
\begin{align}
\mathbb{A}\,\mathbf{n}=\mathbf{K}+\boldsymbol\phi.
\end{align}
\item A generalized gauge transformation 
\begin{align}
n_{i}\to n'_i+\Delta_i,
\end{align}
such that
\begin{align}
\label{fullsyst1}
\mathbb{A}\,\Delta=\boldsymbol\phi,\\
\label{fullsyst2}
\mathbb{A}\,\mathbf{n}'=\mathbf{K},
\end{align}
is performed on the amplitude in order to obtain a new set of numerators satisfying the C/K-duality. The solution of \eqref{fullsyst1} determines the shifts linking the starting set of numerators to the dual representation.
\item The existence of solutions for the the systems \eqref{fullsyst1} and \eqref{fullsyst2} is related to the constraint
\begin{align}
\text{rank}(\mathbb{A}|\boldsymbol\phi)=\text{rank}(\mathbb{A}|\mathbf{K})=\text{rank}(\mathbb{A}).
\label{rank1}
\end{align}
This consistency condition is able to detect all $N-\text{rank}(\mathbb{A})$ non-trivial constraints both between the C/K-violating terms $\phi_{[i,j,k]}$ and the kinematic factors $K_{\sigma(i)}$, the latter corresponding to the well-known  relations between color ordered amplitudes which were first observed, for gluon amplitudes, in ~\cite{Bern:2008qj}.Note the $N-\text{rank}(\mathbb{A})$ also determines the number of completely free parameters the set of C/K-dual numerators will depend on.
\end{itemize}
In the following Section we give an example of this method, determining the C/K-dual representation for $gg\to q\bar{q}g$ and showing that the knowledge of the matrix $\mathbb{A}$ can be used to determine the constraints on kinematic factors $K_{\sigma(i)}$ as well on the anomalies $\phi_{[i,j,k]}$, which rise as direct consequence of the off-shell decompositions worked out in Sections \ref{sec:scalars}-\ref{sec:gluons}.


\section{Color-kinematics duality for  $gg\to q\bar{q}g$}
\label{ex1}
Extensions of C/K duality in QCD amplitudes with fundamental matter have been  discussed in~\cite{Johansson:2015oia,Johansson:2014zca}, where manifest duality has been verified for several processes.\\
In order to illustrate the method proposed in the previous Section, we provide a further example of C/K-duality in QCD, by determining dual numerators for $gg\to q\bar{q}g$.\\
The process under consideration contains a single external quark-antiquark pair and, as a consequence, receives contribution from the four-gluon vertex. This allows us to show, in a concrete case, how contact interactions can be treated. We go step by step through the procedure outlined in Section~\ref{method}, adopting notation and conventions similar to~\cite{Bern:2008qj}.
\begin{figure}[t]
\vspace{0.5cm}
\centering
\includegraphics[scale=0.75]{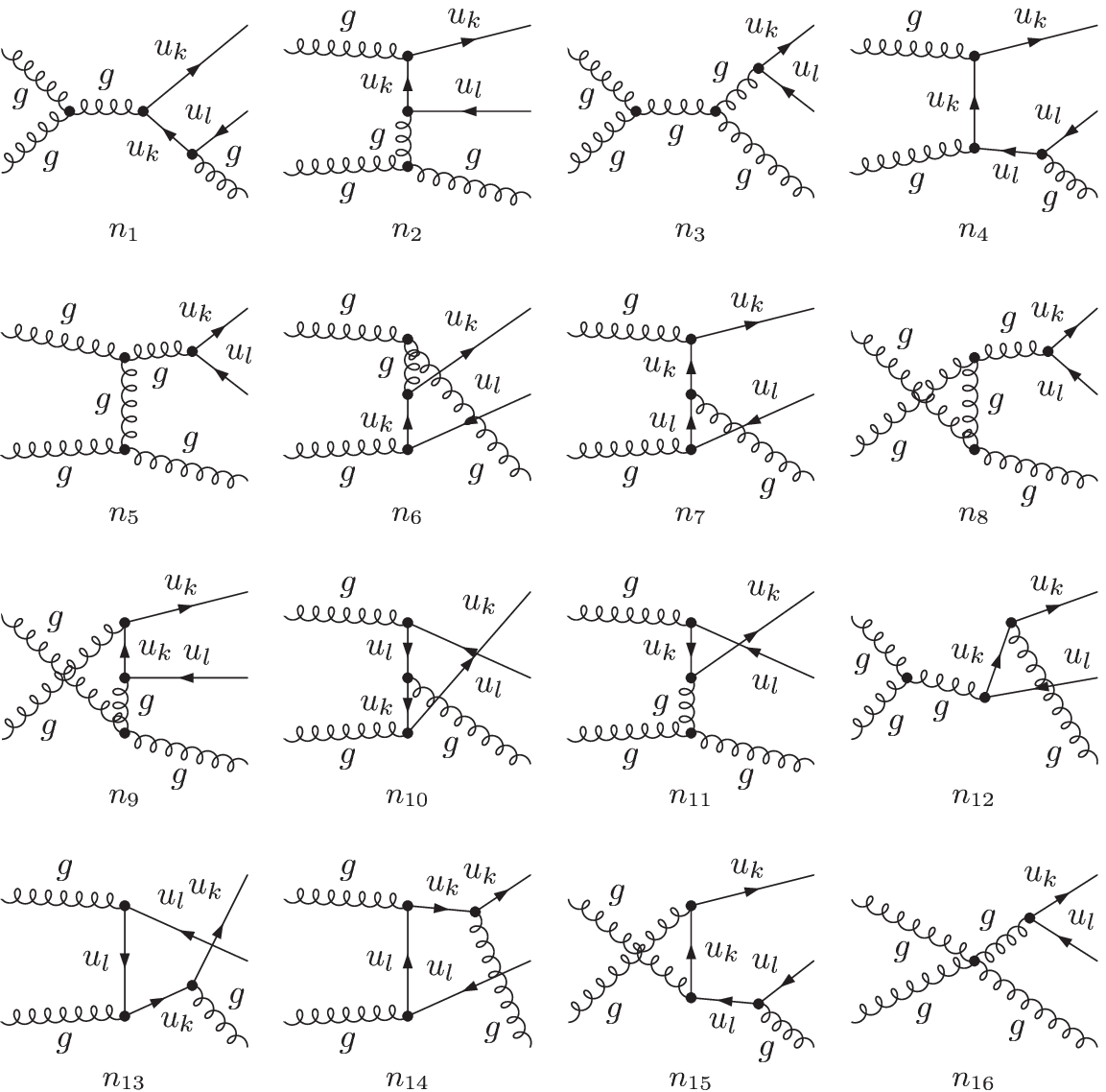}
\caption{Feynman diagrams for $gg\to q\bar q g$}\label{5pt}
\end{figure}

Fig.~\ref{5pt} shows the 16 Feynman diagrams for the process $gg\to q\bar{q}g$. The contribution of $n_{16}$, which, containing the four-gluon vertex, depends on three different color structures, 
\begin{align}
c_{16}n_{16}=c_{3}n_{3;16}+c_{5}n_{5;16}+c_{8}n_{8;16}
\label{4pointdec}
\end{align}
can be split between $n_{3}$, $n_{5}$ and $n_{8}$, so that the new cubic numerators read
\begin{align}
n_{3}+s_{12}n_{3;16}&\to n_{3},\nn
n_{5}+s_{15}n_{5;16}&\to n_{5},\nn
n_{8}+s_{25}n_{3;16}&\to n_{8},
\end{align}
being $s_{ij}=(p_i+p_j)^2$.\\
Thus, the decomposition of the amplitude in terms of cubic graphs reads
\begin{multline}
\mathcal{A}_{5}^{\text{tree}}\big(g(1),g(2),q(3),\bar{q}(4),g(5)\big)=\\[1.5ex]
\frac{c_{1}n_{1}}{s_{12}s_{45}}+\frac{c_{2}n_{2}}{s_{32}s_{51}} +\frac{c_{3}n_{3}}{s_{34}s_{12}}+\frac{c_{4}n_{4}}{s_{45}s_{23}}+\frac{c_{5}n_{5}}{s_{15}s_{34}}\\
+\frac{c_{6}n_{6}}{s_{14}s_{25}}+\frac{c_{7}n_{7}}{s_{23}s_{14}}+\frac{c_{8}n_{8}}{s_{25}s_{34}}+\frac{c_{9}n_{9}}{s_{13}s_{25}}+\frac{c_{10}n_{10}}{s_{24}s_{13}}\\
+\frac{c_{11}n_{11}}{s_{15}s{_{24}}}+\frac{c_{12}n_{12}}{s_{12}s_{35}}+\frac{c_{13}n_{13}}{s_{35}s_{24}}+\frac{c_{14}n_{1}}{s_{14}s_{35}}+\frac{c_{15}n_{15}}{s_{13}s_{45}},
\label{amp5}
\end{multline} 
where only $n_{3}$, $n_{5}$ and $n_{8}$ differ for an additional contact term from the expression given by Feynman rules.\\
 The color factors,
\begin{align}
c_{1} & =\f^{a_{1}a_{2}b}T_{3\bar{k}}^{b}T_{k\bar{4}}^{a_{5}}, &  & c_{2}=T_{3\bar{k}}^{a_{2}}T_{k\bar{4}}^{b}\f^{ba_{5}a_{1}},\nonumber \\
c_{3} & =-T_{3\bar{4}}^{b}\f^{ba_{5}c}\f^{ca_{1}a_{2}}, &  & c_{4}=-T_{3\bar{k}}^{a_{2}}T_{k\bar{l}}^{a_{1}}T_{l\bar{4}}^{a_{5}},\nonumber \\
c_{5} & =-\f^{a_{5}a_{1}b}\f^{ba_{2}c}T_{3\bar{4}}^{c}, &  & c_{6}=T_{3\bar{k}}^{b}T_{k\bar{4}}^{a_{1}}\f^{ba_{2}a_{5}},\nonumber \\
c_{7} & =T_{3\bar{k}}^{a_{2}}T_{k\bar{l}}^{a_{5}}T_{l\bar{4}}^{a_{1}}, &  & c_{8}=-\f^{a_{2}a_{5}b}\f^{ba_{1}c}T_{3\bar{4}}^{c},\nonumber \\
c_{9} & =T_{3\bar{k}}^{a_{1}}T_{k\bar{4}}^{b}\f^{ba_{2}a_{5}}, &  & c_{10}=-T_{3\bar{k}}^{a_{1}}T_{k\bar{l}}^{a_{5}}T_{l\bar{4}}^{a_{2}},\nonumber \\
c_{11} & =\f^{a_{5}a_{1}b}T_{3\bar{k}}^{b}T_{k\bar{4}}^{a_{2}}, &  & c_{12}=T_{3\bar{k}}^{a_{5}}T_{k\bar{4}}^{b}\f^{ba_{1}a_{2}},\nonumber \\
c_{13} & =T_{3\bar{k}}^{a_{5}}T_{k\bar{l}}^{a_{1}}T_{l\bar{4}}^{a_{2}}, &  & c_{14}=-T_{3\bar{k}}^{a_{5}}T_{k\bar{l}}^{a_{2}}T_{l\bar{4}}^{a_{1}},\nonumber \\
c_{15} & =T_{3\bar{k}}^{a_{1}}T_{k\bar{l}}^{a_{2}}T_{l\bar{4}}^{a_{5}}.
\end{align} 
%
%
satisfy a set of 10 Jacobi identities,
\begin{align}
&-c_1+c_3+c_{12}=0,\nn
&-c_1+c_4+c_{15} =0,\nn
&-c_2+c_4+c_7=0,\nn
&-c_2+c_5+c_{11}=0,\nn
&-c_6+c_7+c_{14}=0,\nn
&-c_6+c_8+c_9=0,\nn
&-c_9+c_{10}+c_{15}=0,\nn
&-c_{11}+c_{10}+c_{13}=0,\nn
&-c_{12}+c_{13}+c_{14}=0,\nn
&(-c_5+c_3+c_8=0).
\label{tenjac}
\end{align}

\noindent The system is redundant, since any of the above equations, for instance the last one, can be expressed as a linear combination of the others 9. Therefore, it can be freely dropped.\\
We solve \eqref{tenjac}, choosing $\{c_{1},c_{2},c_{3},c_{4},c_{5},c_{6}\}$ as independent color factor, and we re-express the amplitude as
\begin{align}
\mathcal{A}_{5}^{\text{tree}}\big(g(1),g(2),q(3),\bar{q}(4),g(5)\big)=\sum_{i=1}^{6}c_i K_{i}
\label{jacord}
\end{align}
\begin{align}
 &K_1=\frac{n_{1}}{s_{12}s_{45}}+\frac{n_{12}}{s_{12}s_{35}}+\frac{n_{13}}{s_{24}s_{35}}-\frac{n_{10}}{s_{13}s_{42}}+\frac{n_{15}}{s_{13}s_{45}},\nn
 &K_{2}=\frac{n_{2}}{s_{23}s_{51}}+\frac{n_{7}}{s_{14}s_{32}}-\frac{n_{14}}{s_{14}s_{35}}+\frac{n_{13}}{s_{24}s_{35}}+\frac{n_{11}}{s_{42}s_{51}},\nn
 &K_3=\frac{n_{3}}{s_{12}s_{34}}+\frac{n_{9}}{s_{13}s_{25}}-\frac{n_{12}}{s_{12}s_{35}}-\frac{n_{13}}{s_{24}s_{35}}+\frac{n_{10}}{s_{13}s_{42}}-\frac{n_{8}}{s_{25}s_{43}},\nn
 &K_{4}=\frac{n_{4}}{s_{23}s_{45}}-\frac{n_{7}}{s_{14}s_{32}}+\frac{n_{14}}{s_{14}s_{35}}-\frac{n_{13}}{s_{24}s_{35}}+\frac{n_{10}}{s_{13}s_{42}}-\frac{n_{15}}{s_{13}s_{45}},\nn
 &K_{5}=\frac{n_{5}}{s_{34}s_{51}}-\frac{n_{9}}{s_{13}s_{25}}-\frac{n_{10}}{s_{13}s_{42}}+\frac{n_{8}}{s_{25}s_{43}}-\frac{n_{11}}{s_{42}s_{51}},\nn
 &K_{6}=\frac{n_{6}}{s_{14}s_{25}}+\frac{n_{9}}{s_{13}s_{25}}+\frac{n_{14}}{s_{14}s_{35}}-\frac{n_{13}}{s_{24}s_{35}}+\frac{n_{10}}{s_{13}s_{42}}.
\label{Kfactor}
\end{align}
We remark that this choice is by no means unique and other admissible sets of independent color factors lead to different but equivalent decompositions.\\
\par Now we introduce the shifts \eqref{shifts} and, by imposing generalized gauge invariance on \eqref{jacord}, $\delta\mathcal{A}_{5}^{\text{tree}}=0$, we obtain a set of 6 homogenous equations. Furthermore, in order to establish C/K-duality for the new numerators $n'_i$, we require the shifts to absorb the anomalous terms, \textit{i.e.} to be solution of an additional set of 9 non-homogeneous equations, that are obtained from \eqref{tenjac} by replacing each color factors $c_{i}$ with the corresponding shift $\Delta_{i}$ and the r.h.s. with the proper anomaly.\\
Thus, dual numerators are determined from the solution of the linear system \eqref{system}, where $\mathbb{A}$ is the $15\times 15$ matrix
\newcommand\scalemath[2]{\scalebox{#1}{\mbox{\ensuremath{\displaystyle #2}}}}

\begin{widetext}
\begin{align}
&\mathbb{A}=\nn
&\scalemath{0.875}{\left(
\begin{array}{@{}ccccccccccccccc@{}}
 0 & 0 & 0 & 0 & \frac{1}{s_{34} s_{15}} & 0 & 0 & \frac{1}{s_{25} s_{34}} &
   -\frac{1}{s_{13} s_{25}} & -\frac{1}{s_{13} s_{24}} & -\frac{1}{s_{24} s_{15}} &
   0 & 0 & 0 & 0 \\
 0 & 0 & \frac{1}{s_{12} s_{34}} & 0 & 0 & 0 & 0 & -\frac{1}{s_{25} s_{34}} &
   \frac{1}{s_{13} s_{25}} & \frac{1}{s_{13} s_{24}} & 0 & -\frac{1}{s_{12}
   s_{35}} & -\frac{1}{s_{24} s_{35}} & 0 & 0 \\
 0 & \frac{1}{s_{23} s_{15}} & 0 & 0 & 0 & 0 & \frac{1}{s_{14} s_{23}} & 0 & 0 & 0 &
   \frac{1}{s_{24} s_{15}} & 0 & \frac{1}{s_{24} s_{35}} & -\frac{1}{s_{14}
   s_{35}} & 0 \\
 0 & 0 & 0 & 0 & 0 & \frac{1}{s_{14} s_{25}} & 0 & 0 & \frac{1}{s_{13} s_{25}} &
   \frac{1}{s_{13} s_{24}} & 0 & 0 & -\frac{1}{s_{24} s_{35}} & \frac{1}{s_{14}
   s_{35}} & 0 \\
 0 & 0 & 0 & \frac{1}{s_{23} s_{45}} & 0 & 0 & -\frac{1}{s_{14} s_{23}} & 0 & 0 &
   \frac{1}{s_{13} s_{24}} & 0 & 0 & -\frac{1}{s_{24} s_{35}} & \frac{1}{s_{14}
   s_{35}} & -\frac{1}{s_{13} s_{45}} \\
 \frac{1}{s_{12} s_{45}} & 0 & 0 & 0 & 0 & 0 & 0 & 0 & 0 & -\frac{1}{s_{13} s_{24}} &
   0 & \frac{1}{s_{12} s_{35}} & \frac{1}{s_{24} s_{35}} & 0 & \frac{1}{s_{13}
   s_{45}} \\
      -1 & 0 & 1 & 0 & 0 & 0 & 0 & 0 & 0 & 0 & 0 & 1 & 0 & 0 & 0 \\
 -1 & 0 & 0 & 1 & 0 & 0 & 0 & 0 & 0 & 0 & 0 & 0 & 0 & 0 & 1 \\
 0 & -1 & 0 & 1 & 0 & 0 & 1 & 0 & 0 & 0 & 0 & 0 & 0 & 0 & 0 \\
 0 & -1 & 0 & 0 & 1 & 0 & 0 & 0 & 0 & 0 & 1 & 0 & 0 & 0 & 0 \\
 0 & 0 & 0 & 0 & 0 & -1 & 1 & 0 & 0 & 0 & 0 & 0 & 0 & 1 & 0 \\
 0 & 0 & 0 & 0 & 0 & -1 & 0 & 1 & 1 & 0 & 0 & 0 & 0 & 0 & 0 \\
 0 & 0 & 0 & 0 & 0 & 0 & 0 & 0 & -1 & 1 & 0 & 0 & 0 & 0 & 1 \\
 0 & 0 & 0 & 0 & 0 & 0 & 0 & 0 & 0 & 1 & -1 & 0 & 1 & 0 & 0 \\
 0 & 0 & 0 & 0 & 0 & 0 & 0 & 0 & 0 & 0 & 0 & -1 & 1 & 1 & 0 \\
 \end{array}
\right)}
\label{A15}
\end{align}
and the vector $\boldsymbol\phi$ is given by
\begin{align}
\boldsymbol\phi=(
0,
0,
0,
0,
0,
0,
\phi_{[1,3,12]},
\phi_{[1,4,15]},
\phi_{[2,4,7]},
\phi_{[2,5,11]},
\phi_{[6,7,14]},
\phi_{[6,8,9]},
\phi_{[9,10,15]},
\phi_{[11,10,13]},
\phi_{[12,13,14]})^{T}.
\label{phivec}
\end{align}
\end{widetext}

\noindent The anomalies $\phi_{[i,j,k]}$ can be directly obtained from our off-shell decompositions. We observe that, since we decided to drop, without loss of generality, the last of equations \eqref{tenjac}, which involves a Jacobi identities for gluons, we can express all anomalies in terms of the fermionic current $J_{\text{q}}=J_{\text{q-Feyn}}+J_{\text{q-Ax}}$, whose explicit expression is given by equations \eqref{BCJq} and \eqref{BCJqax}. Likewise, the three anomalies involving the numerators $n_{3}$, $n_{5}$ and $n_{8}$ receive an additional contribution from the four-gluon interaction, since no contact term was considered in the definition of $J_{\text{q}}$,

\begin{widetext}
\begin{align}
\phi_{[1,3,12]} & =\bar{u}_{3}\left[J_{\text{q}}\left(p_{1}+p_{2},p_{4},p_{3},p_{5}\right)_{\alpha\alpha_{5}}\right]v_{4}\Pi^{\alpha\beta}\left(p_{1}+p_{2},q\right)V_{\beta\alpha_{1}\alpha_{2}}\left(-p_{1}-p_{2},p_{1},p_{2}\right)\varepsilon_{1}^{\alpha_{1}}\varepsilon_{2}^{\alpha_{2}}\varepsilon_{5}^{\alpha_{5}}+n_{3;16}=s_{12}\varphi_{[1,3,12]},\nonumber \\
\phi_{[1,4,15]} & =\bar{u}_{3}\left[J_{\text{q}}\left(p_{2},p_{4}+p_{5},p_{3},p_{1}\right)_{\alpha_{2}\alpha_{1}}\left(\slashed p_{4}+\slashed p_{5}\right)\slashed\varepsilon_{5}\right]v_{4}\,\varepsilon_{1}^{\alpha_{1}}\varepsilon_{2}^{\alpha_{2}}=s_{45}\varphi_{[1,4,15]},\nonumber \\
\phi_{[2,4,7]} & =\bar{u}_{3}\left[\slashed\varepsilon_{2}\left(\slashed p_{3}+\slashed p_{2}\right)J_{\text{q}}\left(p_{1},p_{4},p_{3}+p_{2},p_{5}\right)_{\alpha_{1}\alpha_{5}}\right]v_{4}\,\varepsilon_{1}^{\alpha_{1}}\varepsilon_{5}^{\alpha_{5}}=s_{23}\varphi_{[2,4,7]},\nonumber \\
\phi_{[2,5,11]} & =\bar{u}_{3}\left[J_{\text{q}}\left(p_{1}+p_{5},p_{4},p_{3},p_{2}\right)_{\alpha\alpha_{2}}\right]v_{4}\,\Pi^{\alpha\beta}\left(p_{1}+p_{5},q\right)V_{\beta\alpha_{1}\alpha_{5}}\left(-p_{1}-p_{5},p_{1},p_{5}\right)\varepsilon_{1}^{\alpha_{1}}\varepsilon_{2}^{\alpha_{2}}\varepsilon_{5}^{\alpha_{5}}+n_{5;16}=s_{15}\varphi_{[2,5,11]},\nonumber \\
\phi_{[6,7,14]} &= \bar{u}_{3}\left[J_{\text{q}}\left(p_{5},p_{4}+p_{1},p_{3},p_{2}\right)_{\alpha_{5}\alpha_{2}}\left(\slashed p_{4}+\slashed p_{1}\right)\slashed\varepsilon_{1}\right]v_{4}\,\varepsilon_{2}^{\alpha_{2}}\varepsilon_{5}^{\alpha_{5}}=s_{14}\varphi_{[6,7,14]},\nonumber \\
\phi_{[6,8,9]} & =\bar{u}_{3}\left[J_{\text{q}}\left(p_{2}+p_{5},p_{4},p_{3},p_{1}\right)_{\alpha\alpha_{1}}\right]v_{4}\,\Pi^{\alpha\beta}\left(p_{2}+p_{5},q\right)V_{\beta\alpha_{2}\alpha_{5}}\left(-p_{2}-p_{5},p_{2},p_{5}\right)\varepsilon_{1}^{\alpha_{1}}\varepsilon_{2}^{\alpha_{2}}\varepsilon_{5}^{\alpha_{5}}+n_{8;16}=s_{25}\varphi_{[6,8,9]},\nonumber \\
\phi_{[9,10,15]} &
 =\bar{u}_{3}\left[\slashed\varepsilon_{1}\left(\slashed p_{3}+\slashed p_{1}\right)J_{\text{q}}\left(p_{5},p_{4},p_{3}+p_{1},p_{2}\right)_{\alpha_{5}\alpha_{2}}\right]v_{4}\,\varepsilon_{2}^{\alpha_{2}}\varepsilon_{5}^{\alpha_{5}}=s_{13}\varphi_{[9,10,15]},\nonumber \\
\phi_{[11,10,13]} &
 =\bar{u}_{3}\left[J_{\text{q}}\left(p_{1},p_{4}+p_{2},p_{3},p_{5}\right)_{\alpha_{1}\alpha_{5}}\left(\slashed p_{4}+\slashed p_{2}\right)\slashed\varepsilon_{2}\right]v_{4}\,\varepsilon_{1}^{\alpha_{1}}\varepsilon_{5}^{\alpha_{5}}=s_{24}\varphi_{[11,10,13]},\nonumber \\
\phi_{[12,13,14]} &
 =\bar{u}_{3}\left[\slashed\varepsilon_{5}\left(\slashed p_{3}+\slashed p_{5}\right)J_{\text{q}}\left(p_{2},p_{4},p_{3}+p_{5},p_{1}\right)_{\alpha_{2}\alpha_{1}}\right]v_{4}\,\varepsilon_{1}^{\alpha_{1}}\varepsilon_{2}^{\alpha_{2}}=s_{35}\varphi_{[12,13,14]},
 \label{anom}
\end{align}
\end{widetext}

\noindent where $V_{\mu_1\mu_2\mu_3}(p_1,p_2,p_3)$ stands for the kinematic part of three-gluon vertex.\\ We remark that in \eqref{anom} the factorization of Mandelstam invariants $s_{ij}$ follows from the decomposition \eqref{NqP} and that, for the anomalies with an internal gluon propagator, namely $\phi_{[1,3,12]}$, $\phi_{[2,5,11]}$ and $\phi_{[6,8,9]}$, it can be achieved in a straightforward way thanks to the choice of axial gauge.\\
\\
As we have already anticipated, the system of equations is redundant. In fact, by using momentum conservation to express all the invariants $s_{ij}$ in terms of 5 independent ones, for instance $\{s_{12},s_{23},s_{34},s_{45},s_{51}\}$, we obtain
\begin{align}
\text{rank}(\mathbb{A})=11.
\label{rank}
\end{align}
Therefore, if a solution exists, there must be constraints between the non-zero elements of $\boldsymbol\phi$ able to lower the rank of the adjoint matrix. In particular, we expect these relations to correspond to four independent vanishing linear combinations of rows of the matrix $\mathbb{A}$.\\ This observation provides a constructive criterion to find out the constraints between anomalous terms.
\\
First, we build the most general linear combination of rows of the matrix $\mathbb{A}$ and we fix the coefficients by requiring
\begin{align}
\sum_{i=1}^{15}\beta_{i}\mathbb{A}_{i}=0.
\label{lincomb}
\end{align}
According to \eqref{rank}, one can find at most four linear independent solutions $\{\beta_i^{(j)}\}$ to \eqref{lincomb}. \\
Secondly, after selecting an arbitrary complete set of solutions $\{\beta_i^{(j)}\}$, $j=1,2,3,4$, we can verify that
\begin{align}
\sum_{i=1}^{15}\beta_{i}^{(j)}\boldsymbol\phi_{i}=0,\quad \forall j=1,2,3,4,
\label{combphi}
\end{align}
that gives the desired constraints between the C/K-violating terms.\\For this specific case we find,
\begin{align}
&\frac{\phi_{[6,7,14]}}{s_{14}} +\frac{\phi_{[9,10,15]}}{s_{13}}-\frac{\phi_{[12,13,14]}}{s_{35}}-\frac{\phi_{[1,4,15]}}{s_{45}}=0,\nn
&\frac{\phi_{[6,8,9]}}{s_{25}} +\frac{\phi_{[9,10,15]}}{s_{13}}-\frac{\phi_{[12,13,14]}}{s_{35}}-\frac{\phi_{[1,3,12]}}{s_{12}}=0,\nn
&\frac{\phi_{[6,7,14]}}{s_{14}} +\frac{\phi_{[9,10,15]}}{s_{13}}-\frac{\phi_{[11,10,13]}}{s_{24}}-\frac{\phi_{[2,4,7]}}{s_{23}}=0,\nn
&\frac{\phi_{[6,8,9]}}{s_{25}}+\frac{\phi_{[11,10,13]}}{s_{24}}+\frac{\phi_{[2,5,11]}}{s_{15}}-\frac{\phi_{[6,7,14]}}{s_{14}}=0,\nn
\label{relPhi}
\end{align}
which, thanks to \eqref{anom}, can be written as
\begin{align}
    \label{relphi1}
&	\varphi_{[6,7,14]} +\varphi_{[9,10,15]}-\varphi_{[12,13,14]}-\varphi_{[1,4,15]}=0,\\
&	\varphi_{[6,8,9]}+\varphi_{[9,10,15]}-\varphi_{[12,13,14]}-\varphi_{[1,3,12]}=0,\\
	\label{relphi2}
&	\varphi_{[6,7,14]}+\varphi_{[9,10,15]}-\varphi_{[11,10,13]}-\varphi_{[2,4,7]}=0,\\
	\label{relphi3}
&	\varphi_{[6,8,9]}+\varphi_{[11,10,13]}+\varphi_{[2,5,11]}-\varphi_{[6,7,14]}=0.
\end{align}
\\
These relations follow as a direct consequence of the off-shell decomposition \eqref{NqP}. For instance, if we specify it for the anomalies appearing in \eqref{relphi1} by using the explicit form of $J_{\text{q}}$, we obtain\\

\begin{widetext}
\begin{align}
\varphi_{[1,4,15]}&=\bar{u}_3\bigg[\gamma^{\alpha_2}\gamma^{\alpha_1}\gamma^{\alpha_5}-g^{\alpha_{1}\alpha_2}\gamma^{\alpha_5}-\frac{g^{\alpha_1\alpha_2}q\cdot (p_1-p_2)-2(q^{\alpha_1}p_1^{\alpha_2}-q^{\alpha_2}p_2^{\alpha_1})}{q\cdot(p_1+p_2)}\gamma^{\alpha_5}\bigg]v_4\varepsilon_{\alpha_1}\varepsilon_{\alpha_2}\varepsilon_{\alpha_5},\nn
\varphi_{[6,7,14]}&=\bar{u}_3\bigg[\gamma^{\alpha_5}\gamma^{\alpha_2}\gamma^{\alpha_1}-g^{\alpha_{2}\alpha_5}\gamma^{\alpha_1}-\frac{g^{\alpha_2\alpha_5}q\cdot (p_2-p_5)-2(q^{\alpha_2}p_2^{\alpha_5}-q^{\alpha_5}p_5^{\alpha_2})}{q\cdot(p_2+p_5)}\gamma^{\alpha_1}\bigg]v_4\varepsilon_{\alpha_1}\varepsilon_{\alpha_2}\varepsilon_{\alpha_5},\nn
\varphi_{[12,13,14]}&=\bar{u}_3\bigg[-\gamma^{\alpha_5}\gamma^{\alpha_1}\gamma^{\alpha_2}+g^{\alpha_{1}\alpha_2}\gamma^{\alpha_5}+\frac{g^{\alpha_1\alpha_2}q\cdot (p_1-p_2)-2(q^{\alpha_1}p_1^{\alpha_2}-q^{\alpha_2}p_2^{\alpha_1})}{q\cdot(p_1+p_2)}\gamma^{\alpha_5}\bigg]v_4\varepsilon_{\alpha_1}\varepsilon_{\alpha_2}\varepsilon_{\alpha_5},\nn
\varphi_{[9,10,15]}&=\bar{u}_3\bigg[-\gamma^{\alpha_1}\gamma^{\alpha_2}\gamma^{\alpha_5}+g^{\alpha_{2}\alpha_5}\gamma^{\alpha_1}+\frac{g^{\alpha_2\alpha_5}q\cdot (p_2-p_5)-2(q^{\alpha_2}p_2^{\alpha_5}-q^{\alpha_5}p_5^{\alpha_2})}{q\cdot(p_2+p_5)}\gamma^{\alpha_1}\bigg]v_4\varepsilon_{\alpha_1}\varepsilon_{\alpha_2}\varepsilon_{\alpha_5}.
\label{deco}
\end{align}
\end{widetext}

By using Clifford algebra we can verify that
\begin{align}
\varphi_{[1,4,15]}+\varphi_{[12,13,14]}-\varphi_{[6,7,14]} -\varphi_{[9,10,15]}&=\nn
2\varepsilon_{\alpha_1}\varepsilon_{\alpha_2}\varepsilon_{\alpha_5}\bar{u}_3\big[g^{\alpha_1\alpha_2}\gamma^{\alpha_5}-g^{\alpha_1\alpha_2}\gamma^{\alpha_5}\big]v_4&=0.
\end{align}
Similar cancellation are encountered in all other cases.\\
The constraints \eqref{relPhi} make the consistency relation satisfied,
\begin{align}
\text{rank}(\mathbb{A}|\phi)=\text{rank}(\mathbb{A})=11.
\end{align}
 As a consequence, the system admits a solution which leaves four shifts completely undetermined and the amplitude has a C/K-dual representation, consistent with generalized gauge invariance, whose numerators depend of four free parameters. This number agrees with the $(n-2)!-(n-3)!$ degrees of freedom found in ~\cite{Bern:2008qj} for the pure Yang-Mills case.
\\
\par In order to find an explicit expressions for the shifts, we  build a maximum-rank system by selecting a subset of 11 independent equations and proceed by Gaussian elimination. We observe that, as for the solution of the Jacobi identities for color factors, also in this case there is a remarkably large freedom in the choice of the independent equations to be solved and, furthermore, in the set of four arbitrary shifts to appear in the solution.\\ 
In our case, by selecting equations corresponding to rows 1-2 and 9-15 of \eqref{A15}, we express $\Delta_{i}$, $i=5,6,...,15$ as linear combination of the anomalies $\phi_{[i,j,k]}$ and of the four arbitrary shifts $\{\Delta_{1},\Delta_{2},\Delta_{3},\Delta_{4}\}$,
\begin{align}
\Delta_{i}=\sum_{j=1}^{4}\mathcal{R}_{ij}(s_{kl})\Delta_{j}+&\sum_{[m,n,p]}\mathcal{R}_{i[m,n,p]}(s_{kl})\phi_{[m,n,p]},\nn
&\qquad\qquad\quad\text{for }i=5,...15,\label{solution}
\end{align}
where $\mathcal{R}_{ij}$ and $\mathcal{R}_{i[mnp]}$ are dimensionless rational functions of the invariants $s_{kl}$.\\ 
The analytic expression of \eqref{solution}, which is not provided here for sake of simplicity, has been obtained for arbitrary polarizations and has been numerically checked for all helicity configurations. In particular, the complete independence on the actual values of the four independent shifts has been verified for the full color-dressed amplitude as well as for each ordering appearing in \eqref{jacord}. We observe that the choice $\Delta_{i}=0$, $i=1,...,4$ leads to a dual representation where four numerators correspond exactly to the starting ones and three anomalous terms are attributed to single diagrams
\begin{align}
\Delta_{7}&=\phi_{[2,4,7]},\nn
\Delta_{12}&=\phi_{[1,3,12]},\nn
\Delta_{15}&=\phi_{[1,4,15]}.
\end{align}
In addition, for any choice of the free parameters, the set of new numerators $n'_i$, satisfies the system of equations \eqref{systemnh}, where
\begin{align}
\mathbf{K}=(K_1,K_2,...,K_6,0,0,...,0)^{T}.
\end{align}
Therefore, the consistency requirement
\begin{align}
\text{rank}(\mathbb{A}|\mathbf{K})=\text{rank}(\mathbb{A})=11
\end{align}
allow us to use the exactly the same solutions $\{\beta_{i}^{(j)}\}$ of \eqref{lincomb}, to establish relations between the kinematic factors,
\begin{align}
\sum_{i=1}^{15}\beta_{i}^{(j)}\mathbf{K}_i=0,\quad \forall j=1,2,...,4,
\label{combK}
\end{align}
which, in this case, read
\begin{align}
&s_{45}K_1-s_{34}K_3-s_{14}K_6=0,\nn
&s_{12}K_1-s_{23}K_4-s_{25}K_6=0,\nn
&s_{15}K_2-s_{45}K_4-s_{25}K_6=0,\nn
&s_{23}K_2-s_{34}K_5+(s_{23}+s_{35})K_6=0.
\label{relK}
\end{align}
This set of constraints reduces from 6 to 2 the number of independent kinematic factors.\\
Therefore, as we have already pointed out, \eqref{relK} can be considered as equivalent to the well-known monodromy relations which have been shown, for the pure-gluon case, to reduce to $(n-3)!$ the number of independent color ordered amplitudes.\\
 Nevertheless, we want to remark that, in the approach we have presented, the origin of \eqref{relK}, as well as the one of \eqref{relPhi}, is shown to be purely diagrammatic. In particular, whereas in ~\cite{BjerrumBohr:2010zs}  monodromy relations analogous to \eqref{relK} are derived from the field-limit of string theory and a set of relations equivalent to \eqref{relPhi} is presented as a parametrization their solution, here both are derived as a necessary consequence of the redundancy of kinematic matrix $\mathbb{A}$ and they have be shown to naturally emerge from the off-shell decomposition of the Jacobi-combination of kinematic numerators in axial gauge.


\section{Color-kinematics duality in $d$-dimensions}
\label{FDFduality}
In this Section we study the C/K-duality for tree-level amplitudes in
dimensional regularization. They are the basic building
blocks for the determination of higher-order scattering amplitudes within generalized unitarity based methods.
We employ the four dimensional formulation (FDF) scheme, recently
introduced in~\cite{Fazio:2014xea}. Within FDF, the additional degrees of
freedom which naturally enters when the space-time dimensions are
continued beyond four, such as spinors and polarizations, admit
a purely four-dimensional representation.
FDF has been successfully applied to reproduce one-loop corrections to
$gg \to gg$, $q {\bar q} \to gg$, $gg \to Hg$ (in the heavy top
limit), as well as $gg \to gggg$ \cite{Bobadilla:2015wma}.
Accordingly, the states propagating around the loop are described as
four dimensional massive particles. The four-dimensional degrees
of freedom of the gauge bosons are carried by  {\it massive vector bosons} (denoted by $g^{\bullet}$)
of mass $\mu$ (associated to three polarization states) and their
$(d-4)$-dimensional ones by {\it real scalar particles} ($s^{\bullet}$) of mass
$\mu$, being $\mu$ an extra-dimensional mass-like parameter. A
$d$-dimensional fermion
of mass $m$ is instead traded for a {\it tardyonic  Dirac field} ($q^{\bullet}$) with
mass $m +i \mu \gamma^5$~\cite{Jentschura:2012vp} (associated to two
spinor states). The $d$ dimensional algebraic manipulations are
replaced by four-dimensional ones complemented by
a set of multiplicative selection rules. 
The latter are treated as an algebra describing internal symmetries. 
In \ref{AppA}, for completeness, we
provide the Feynman rules of the FDF scheme.

We anticipate that the C/K duality obeyed by the numerators of
tree-level amplitudes within the FDF scheme are non-trivial relations involving the
interplay of massless and massive particles.


 \subsection{Tree-level identities in $d$-dimensions}
 \label{dtree}
As a starting point, we focus on four-point tree-level amplitudes, showing that C/K-duality can be established for all interactions involving particles propagating within the FDF framework.\\
Envisaging one-loop applications of these results, we consider amplitudes with two $d$-dimensional external particles and two four-dimensional ones. Following the by now familiar procedure of Sections \ref{sec:scalars}, \ref{sec:quarks} and \ref{sec:gluons}, we build an off-shell Jacobi-like combination of kinematics numerators for each of the seven processes involving, according to the Feynman rules of \ref{AppA}, two external massive particles. We show that, in every case, C/K-duality holds after physical constraints are imposed.\\
\begin{enumerate}[a)]
\item We start from the scattering of two generalized gluons producing two massless ones, $g^{\bullet}g^{\bullet}\to gg$, with on-shell conditions $p_2^2=p_3^2=0$ and $p_2^2=p_3^2=\mu^2$. The amplitude receives contribution from the four diagrams shown in Fig.\ref{4diagFDF}, where dotted lines indicate generalized particles.
	\begin{figure}[t]
		\centering
		\vspace*{0.5cm}
		\includegraphics[scale=1.0]{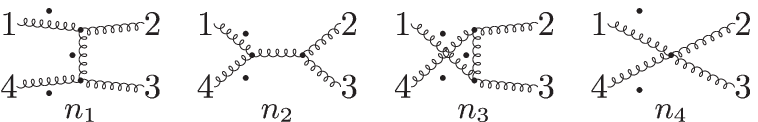}
		\caption{Feynman diagrams for $g^{\bullet}g^{\bullet} \to gg$.}
		\label{4diagFDF}
	\end{figure}
Exactly as in the massless case discussed in Section \ref{sec:gluons}, the four-point vertex contributes to the same color structures as the trivalent diagrams. Hence it can be decomposed as 
\begin{align}
c_{4}n_{4}=c_{1}n_{1;4}+c_{2}n_{2;4}+c_{3}c_{3;4},
\end{align}
so that, by absorbing its kinematic part into the numerators of the first three diagrams, the amplitude can be expressed in terms of cubic graphs only.\\
We observe that $n_2$ is associated to the massless denominator $(p_1+p_4)^2$ while $n_1$ and $n_3$ sit, respectively, on the massive propagators $\big((p_1+p_2)^2-\mu^2\big)$ and $\big((p_1+p_3)^2-\mu^2\big)$.\\
Therefore, the proper definition of the cubic numerators absorbing the contact interaction is
\begin{align}
&n_{1}+((p_1+p_2)^2-\mu^2\big)n_{1;4}\to n_{1} ,\nn &n_{2}+(p_1+p_4)^2n_{2;4}\to n_{2},\nn
&n_{3}+((p_1+p_3)^2-\mu^2\big)n_{3;4}\to n_{3}.
\end{align}
With this prescription, the Jacobi-combination of the kinematic numerators depicted in Fig.\ref{BCJFDFa} can be written as
\begin{align}
N_{\text{g}^{\bullet}\text{g}^{\bullet}\text{g}\text{g}}=J_{\text{g}^{\bullet}\text{g}^{\bullet}\text{gg}}^{\mu_{1}...\mu_{4}} \varepsilon_{\mu_1}(p_1) \varepsilon_{\mu_2}(p_2) \varepsilon_{\mu_3}(p_3)\varepsilon_{\mu_4}(p_4),
\end{align} 
with
\begin{align}
J_{\text{g}^{\bullet}\text{g}^{\bullet}\text{gg}}^{\mu_{1}\mu_{2}\mu_{3}\mu_{4}}=&J_{\text{g-Fey}}^{\mu_{1}\mu_{2}\mu_{3}\mu_{4}}+J_{\text{g}^{\bullet}\text{g}^{\bullet}\text{gg}\text{-Ax}}^{\mu_{1}\mu_{2}\mu_{3}\mu_{4}}+J_{\text{g}^{\bullet}\text{g}^{\bullet}\text{gg}\text{-$\mu^2$}}^{\mu_{1}\mu_{2}\mu_{3}\mu_{4}}.
\label{J-GGgg}
\end{align}
In Eq.~\eqref{J-GGgg} $J_{\text{g-Fey}}$ is the same as in \eqref{jfeyg}; $J_{\text{g}^{\bullet}\text{g}^{\bullet}\text{gg}\text{-Ax}}$ is given by
\begin{multline}
-i\,J_{\text{g}^{\bullet}\text{g}^{\bullet}\text{gg}\text{-Ax}}=\frac{1}{\left(p_{2}+p_{3}\right)\cdot q}\Big\{\\
\left(p_{4}^{\mu_{1}}p_{4}^{\mu_{4}}-p_{1}^{\mu_{1}}p_{1}^{\mu_{4}}+\left(p_{1}^{2}-p_{4}^{2}\right)g^{\mu_{1}\mu_{4}}\right)[q\cdot\left(p_{2}-p_{3}\right)g^{\mu_{2}\mu_{3}}\\
+\left(2p_{3}+p_{2}\right)^{\mu_{2}}q^{\mu_{3}}-\left(2p_{2}+p_{3}\right)^{\mu_{3}}q^{\mu_{2}}]\\
+\left(p_{3}^{\mu_{2}}p_{3}^{\mu_{3}}-p_{2}^{\mu_{2}}p_{2}^{\mu_{3}}+\left(p_{2}^{2}-p_{3}^{2}\right)g^{\mu_{2}\mu_{3}}\right)\big[q\cdot\left(p_{4}-p_{1}\right)g^{\mu_{1}\mu_{4}}\\
+\left(2p_{1}+p_{4}\right)^{\mu_{4}}q^{\mu_{1}}-\left(2p_{4}+p_{1}\right)^{\mu_{1}}q^{\mu_{4}}]\Big\},
\label{J-GGggax}
\end{multline}
whereas the $\mu$-dependent term reads
\begin{multline}
-i\, J_{\text{g}^{\bullet}\text{g}^{\bullet}\text{gg}\text{-$\mu^2$}}^{\mu_{1}\mu_{2}\mu_{3}\mu_{4}}=\frac{1}{\mu^2}\Big\{\\ g^{\mu_{1}\mu_{3}}g^{\mu_{2}\mu_{4}}\left(\left(p_{1}^{2}-p_{3}^{2}\right)\left(p_{4}^{2}-p_{2}^{2}\right)-\mu^{4}\right)\\-g^{\mu_{1}\mu_{2}}g^{\mu_{3}\mu_{4}}\left(\left(p_{1}^{2}-p_{2}^{2}\right)\left(p_{4}^{2}-p_{3}^{2}\right)-\mu^{4}\right)\\-\left(p_{4}^{2}-p_{2}^{2}\right)g^{\mu_{2}\mu_{4}}\left(p_{1}^{\mu_{1}}p_{1}^{\mu_{3}}-p_{3}^{\mu_{1}}p_{3}^{\mu_{3}}\right)\\+\left(p_{1}^{2}-p_{3}^{2}\right)g^{\mu_{1}\mu_{3}}\left(p_{2}^{\mu_{2}}p_{2}^{\mu_{4}}-p_{4}^{\mu_{2}}p_{4}^{\mu_{4}}\right)\\-\left(p_{1}^{2}-p_{2}^{2}\right)g^{\mu_{1}\mu_{2}}\left(p_{3}^{\mu_{3}}p_{3}^{\mu_{4}}-p_{4}^{\mu_{3}}p_{4}^{\mu_{4}}\right)\\+\left(p_{4}^{2}-p_{3}^{2}\right)g^{\mu_{3}\mu_{4}}\left(p_{1}^{\mu_{1}}p_{1}^{\mu_{2}}-p_{2}^{\mu_{1}}p_{2}^{\mu_{2}}\right)\\+\left(p_{3}^{\mu_{1}}p_{3}^{\mu_{3}}-p_{1}^{\mu_{1}}p_{1}^{\mu_{3}}\right)\left(p_{2}^{\mu_{2}}p_{2}^{\mu_{4}}-p_{4}^{\mu_{2}}p_{4}^{\mu_{4}}\right)\\+\left(p_{2}^{\mu_{1}}p_{2}^{\mu_{2}}-p_{1}^{\mu_{1}}p_{1}^{\mu_{2}}\right)\left(p_{4}^{\mu_{3}}p_{4}^{\mu_{4}}-p_{3}^{\mu_{3}}p_{3}^{\mu_{4}}\right)\Big\}.
\label{J-GGggmu}
\end{multline}
By inspection of ~\eqref{J-GGggmu} and ~\eqref{J-GGggax} we see that C/K duality, which corresponds to $N_{\text{g}^{\bullet}\text{g}^{\bullet}\text{g}\text{g}}=0$, is recovered because of transversality, $\varepsilon_{i}\cdot p_{i}=0$, and on-shellness conditions.
\begin{figure}[t]
	\centering
	\vspace*{0.5cm}
	\includegraphics[scale=1.05]{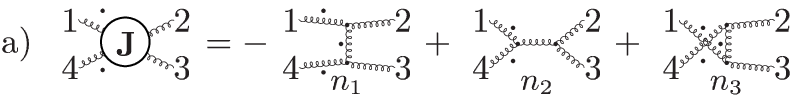}
	\caption{Jacobi combination for $g^{\bullet}g^{\bullet} \to gg$.}
	\label{BCJFDFa}
\end{figure}
\item Now we consider $gg\to s^{\bullet}s^{\bullet}$ ($p_1^2=p_4^2=0$, $p_2^2=p_3^2=\mu^2$) amplitude,
 whose diagrams are shown in Fig.\ref{4diagFDFs}.\\
 In this case, since the four-point interaction only contributes to two color structures,
  \begin{align}
 c_{4}n_{4}=c_{1}n_{1;4}+c_{3}c_{3;4},
 \end{align}
 we can absorb its kinematic part in the two diagrams involving a massive scalar propagator, through the substitutions,
 \begin{align}
 &n_{1}+((p_1+p_2)^2-\mu^2\big)n_{1;4}\to n_{1} ,\nn 
 &n_{3}+((p_1+p_3)^2-\mu^2\big)n_{3;4}\to n_{3},
 \end{align}
 whereas $n_2$ stays the same as defined by Feynman rules.
 In this way, the Jacobi combination of the cubic numerators, depicted in Fig.\ref{BCJFDFb}, becomes
 \begin{align}
 N_{\text{gg}\text{s}^{\bullet}\text{s}^{\bullet}}=J_{\text{gg}\text{s}^{\bullet}\text{s}^{\bullet}}^{\mu_{1}\mu_{4}} \varepsilon_{\mu_1}(p_1)\varepsilon_{\mu_4}(p_4),
 \end{align} 
 where the off-shell current is
  \begin{align}
 J_{\text{gg}\text{s}^{\bullet}\text{s}^{\bullet}}^{\mu_{1}\mu_{4}}=&\big(J_{\text{g-Fey}}^{\mu_{1}\mu_{4}}+J_{\text{g-Ax}}^{\mu_{1}\mu_{4}}\big)G^{AB}.
 \end{align}
 We observe that, although $n_1$ and $n_3$ depend on the mass of the scalar particle, the $\mu$-dependence cancels in their combination. Therefore, C/K-duality, \textit{i.e} $N_{\text{gg}\text{s}^{\bullet}\text{s}^{\bullet}}=0$, holds in this case as well as in the massless one, according to Eqs.~\eqref{jsfey} and \eqref{jsax}.\\
  \begin{figure}[t]
  	\centering
  	\vspace*{0.5cm}
  	\includegraphics[scale=1.0]{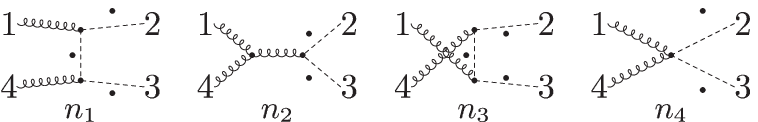}
  	\caption{Feynman diagrams for $gg\to s^{\bullet}s^{\bullet}$.}
  	\label{4diagFDFs}
  \end{figure}
   \begin{figure}[h]
  	\centering
  	\vspace*{0.5cm}
  	\includegraphics[scale=1.05]{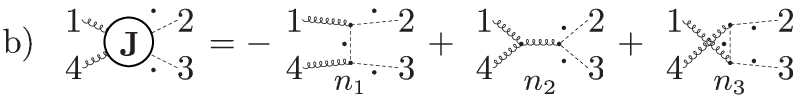}
  	\caption{Jacobi combination for $gg\to s^{\bullet}s^{\bullet}$.}
  	\label{BCJFDFb}
  \end{figure}
\end{enumerate}

\begin{figure}[htb]
		\vspace*{0.5cm}
		\includegraphics[scale=1.05]{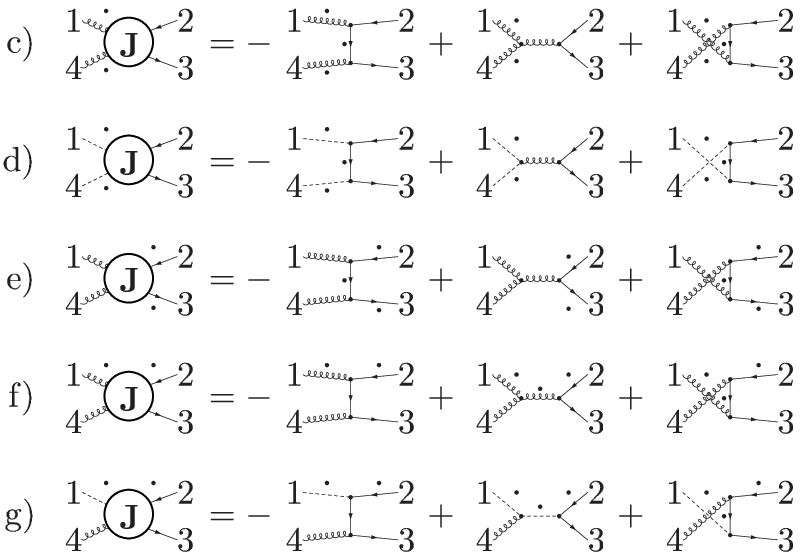}
		\caption{Jacobi combinations for FDF particles.}
		\label{BCJFDF}
	\end{figure}

The remaining processes, whose tree-level identities are depicted in
Fig.\ref{BCJFDF}.(c)-(g) do not involve contact interactions, so that
the construction of their Jacobi-combinations directly follows from
the Feynman rules of \ref{AppA}. In order to avoid repetitive
discussion, we simply list the results giving, for each process, the
corresponding on-shell conditions and the expression of the
Jacobi-combination in terms of off-shell currents. \\
According to the
case, C/K-duality is recovered once transversality of the gluon
polarizations and Dirac equation are taken into account. 
We recall that, for a generalized gluon $g^{\bullet}$ of momentum $p$,
one has $\varepsilon_i \cdot p =0$ $(i=\pm,0)$, while for 
generalized quarks  $q^{\bullet}$ of momentum $p$, one has  
$\bar{u}(p)(\slashed{p}-i\mu\gamma^{5})=0$ and 
	$(\slashed{p}+i\mu\gamma^5)v(p)=0$.

\begin{enumerate}[a)]
	\setcounter{enumi}{2}
	\item $g^{\bullet}g^{\bullet}\to \bar{q}q$ ($p_1^2=p_4^2=\mu^2$, $p_2^2=p_3^2=0$):
	\begin{align}
	N_{\text{g}^{\bullet}\text{g}^{\bullet}\bar{\text{q}}\text{q}}=\bar{u}(p_3)J_{\text{g}^{\bullet}\text{g}^{\bullet}\bar{\text{q}}\text{q}}^{\mu_{1}\mu_{4}}v(p_2)\varepsilon_{\mu_1}(p_1)\varepsilon_{\mu_4}(p_4),
	\end{align}
	with
	\begin{align}
	-J_{\text{g}^{\bullet}\text{g}^{\bullet}\bar{\text{q}}\text{q}}^{\mu_{1}\mu_{4}}=J_{\text{q-Fey}}^{\mu_{1}\mu_{4}}+J_{\text{q-Ax}}^{\mu_{1}\mu_{4}},
	\end{align}
	where the terms in the r.h.s. are defined by Eq.~\eqref{BCJq} and \eqref{BCJqax}.\\
   \item $s^{\bullet}s^{\bullet}\to \bar{q}q$ ($p_1^2=p_4^2=\mu^2$, $p_2^2=p_3^2=0$):
   \begin{align}
   N_{\text{s}^{\bullet}\text{s}^{\bullet}\bar{\text{q}}\text{q}}=\bar{u}(p_3)J_{\text{s}^{\bullet}\text{s}^{\bullet}\bar{\text{q}}\text{q}}v(p_2),
   \end{align}
   with
   \begin{multline}
	-i\,J_{\text{s}^{\bullet}\text{s}^{\bullet}\bar{\text{q}}\text{q}}=\frac{1}{\left(p_{1}+p_{4}\right)\cdot q}\Big\{\\q\cdot\left(p_{4}-p_{1}\right)\left(\slashed p_{2}+\slashed p_{3}\right)-\left(p_{4}^{2}-p_{1}^{2}\right)\slashed q\Big\}.
   \label{zero}
	\end{multline}
	\item $gg\to \bar{q}^{\bullet}q^{\bullet}$ ($p_1^2=p_4^2=0$, $p_2^2=p_3^2=\mu^2$):
	 \begin{align}
	 N_{\text{gg}\bar{\text{q}}^{\bullet}\text{q}^{\bullet}}=\bar{u}(p_3)J_{\text{gg}\bar{\text{q}}^{\bullet}\text{q}^{\bullet}}^{\mu_{1}\mu_{4}}v(p_2),
	 \end{align}
	 with
	\begin{multline}
	-i\,J_{\text{gg}\bar{\text{q}}^{\bullet}\text{q}^{\bullet}}^{\mu_{1}\mu_{4}}=-\left(\slashed p_{3}-i\mu\gamma^{5}\right)\gamma^{\mu_{4}}\gamma^{\mu_{1}}\\
-\gamma^{\mu_{1}}\gamma^{\mu_{4}}\left(\slashed p_{2}+i\mu\gamma^{5}\right)+\left(\slashed p_{3}+\slashed p_{2}+i\mu\gamma^{5}\right)g^{\mu_{1}\mu_{4}}\\
+p_{4}^{\mu_{4}}\gamma^{\mu_{1}}-p_{1}^{\mu_{1}}\gamma^{\mu_{4}}+J_{\text{q-Ax}}^{\mu_{1}\mu_{4}}.
	\end{multline}
	\item $g^{\bullet}g\to \bar{q}^{\bullet}q$ ($p_1^2=p_2^2=\mu^2$, $p_3^2=p_4^2=0$):
	 \begin{align}
	 N_{\text{g}^{\bullet}\text{g}\bar{\text{q}}^{\bullet}\text{q}}=\bar{u}(p_3)J^{\mu_{1}\mu_{4}}_{\text{g}^{\bullet}\text{g}\bar{\text{q}}^{\bullet}\text{q}}v(p_2)\varepsilon_{\mu_1}(p_1)\varepsilon_{\mu_4}(p_4),
	 \end{align}
	 with
	\begin{multline}	-i\,J^{\mu_{1}\mu_{4}}_{\text{g}^{\bullet}\text{g}\bar{\text{q}}^{\bullet}\text{q}}=-\slashed p_{3}\gamma^{\mu_{4}}\gamma^{\mu_{1}}-\gamma^{\mu_{1}}\gamma^{\mu_{4}}\left(\slashed p_{2}+i\mu\gamma^{5}\right)\\+\left(\slashed p_{3}+\slashed p_{2}+i\mu\gamma^{5}\right)g^{\mu_{1}\mu_{4}}+p_{4}^{\mu_{4}}\gamma^{\mu_{1}}-p_{1}^{\mu_{1}}\gamma^{\mu_{4}}\\+\frac{1}{\mu^{2}}\left(\slashed p_{3}+\slashed p_{2}+i\mu\gamma^{5}\right)\left(p_{1}^{\mu_{4}}p_{1}^{\mu_{1}}-p_{4}^{\mu_{1}}p_{4}^{\mu_{4}}\right).
	\end{multline}
	\item $s^{\bullet}g\to \bar{q}^{\bullet}q$ ($p_1^2=p_2^2=\mu^2$, $p_3^2=p_4^2=0$):
	 \begin{align}
	 N_{\text{s}^{\bullet}\text{g}\bar{\text{q}}^{\bullet}\text{q}}=\bar{u}(p_3)J^{\mu_{4}}_{\text{s}^{\bullet}\text{g}\bar{\text{q}}^{\bullet}\text{q}}v(p_2)\varepsilon_{\mu_4}(p_4),
	 \end{align}
	 with
	\begin{equation}
	-i\, J^{\mu_{4}}_{\text{s}^{\bullet}\text{g}\bar{\text{q}}^{\bullet}\text{q}}=\bigg(-\slashed p_{3}\gamma^{\mu_{4}}\gamma^{5}-\gamma^{5}\gamma^{\mu_{4}}\left(\slashed p_{2}+i\mu\gamma^{5}\right)+p_{4}^{\mu_{4}}\gamma^{5}\bigg)\Gamma^{A}.
	\end{equation}

\end{enumerate}

	In summary, we have shown that C/K-duality can be extended to
        four-point tree-level amplitudes in FDF, which would naturally enter the
        construction of loop-level amplitudes in the framework of
        $d$-dimensional unitarity. This constitute the second main
        result of this communication.
	In the concluding part of this letter, as a further application of the method described in Section \ref{method}, we will derive a diagrammatic C/K-dual representation for a five-point building-block involving generalized fields.

\subsection{Color-kinematics duality for $g^{\bullet}g^{\bullet}(s^{\bullet}s^{\bullet})\to q\bar{q}g$}

\begin{figure}[t]
	\vspace{0.5cm}
	\centering
	\includegraphics[scale=0.7]{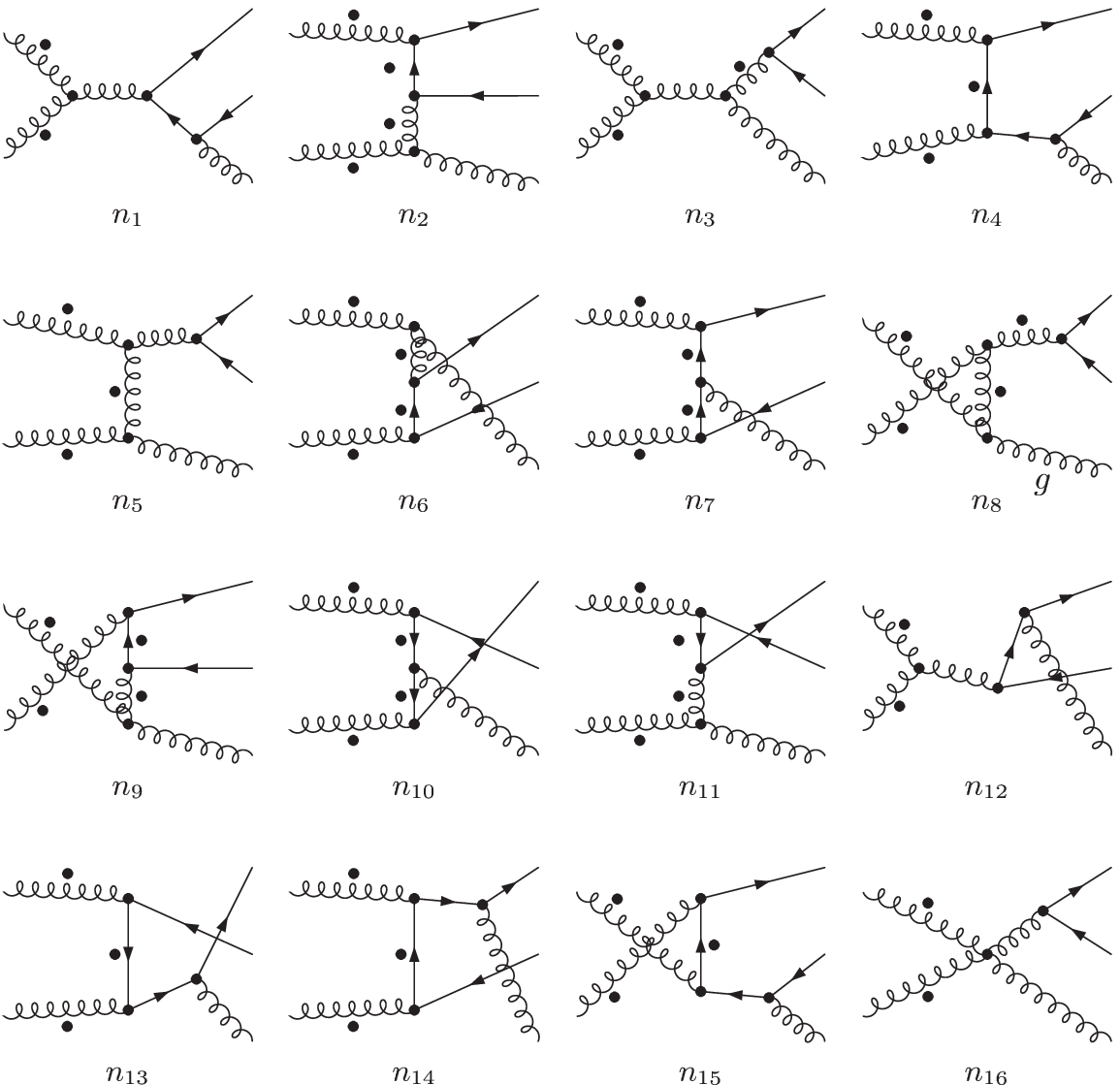}
	\caption{Feynman diagrams for $g^{\bullet}g^{\bullet}\to q\bar q g$. The contributions to $s^{\bullet}s^{\bullet}\to q\bar q g$ are obtained by replacing all generalized gluons $g^{\bullet}$ with $s^\bullet$ lines. }\label{5ptFDF}
\end{figure}

As a non-trivial example of C/K-duality for dimensionally regulated amplitudes, we consider again the process $gg\to q\bar{q}g$, already discussed in Section \ref{ex1}, but now regarding the initial state gluons as $d$-dimensional particles (whereas the final state remains fully four-dimensional). This amplitude would contribute to a $d$-dimensional unitarity cut of a loop-level amplitude where the gluons $p_1$ and $p_2$ appear as virtual states.\\
Within FDF, the full amplitude is obtained by combining the contributions of three different processes involving generalized four-dimensional initial particles,
\begin{align}
&g^{\bullet}g^{\bullet}\to q\bar{q}g,\nn
&s^{\bullet}s^{\bullet}\to q\bar{q}g,\nn
&g^{\bullet}s^{\bullet}\to q\bar{q}g.
\end{align}
However, thanks to the selection rules illustrated in \ref{AppA},  $g^{\bullet}s^{\bullet}\to q\bar{q}g$ vanishes and the problem decouples in the determination of the C/K-dual representation of two individually gauge invariant amplitudes, $g^{\bullet}g^{\bullet}\to q\bar{q}g$ and $s^{\bullet}s^{\bullet}\to q\bar{q}g$. The tree-level contributions to $g^{\bullet}g^{\bullet}\to q\bar{q}g$ are shown in Fig.\ref{5ptFDF}. The Feynman diagrams for $s^{\bullet}s^{\bullet}\to q\bar{q}g$ can be easily obtained by replacing all generalized gluons lines $g^{\bullet}$ with scalars $s^{\bullet}$.
\\Since in both cases the number of graphs, the relations among their color factors and, as consequence, the set of constraints to be imposed on the shifted numerators \eqref{shifts} are exactly the same as in the example of Section \ref{ex1}, we will simply discuss the relevant modifications to be taken into account in order to adapt the calculation to generalized fields.\\

We notice that, while for $g^{\bullet}g^{\bullet}$ the redistribution of numerator $n_{16}$ among cubic diagrams is still given by Eq.~\eqref{4pointdec}, for the $s^{\bullet}s^{\bullet}$ case we have, as discussed in Section \ref{dtree}(b),
\begin{align}
c_{16}n_{16}=c_{5}n_{5;16}+c_{8}n_{8;16}.
\end{align}
\\
Nevertheless, for both processes the diagrammatic expansion of the amplitude can be still read from the r.h.s. of Eq.~\eqref{amp5}, provided the replacement
\begin{align}
&s_{1i}\to (s_{1i}-\mu^2)\;\text{for}\; i\neq 2,\nn
&s_{2i}\to (s_{2i}-\mu^2)\;\text{for}\; i\neq 1,
\end{align}
which account for internal massive propagators. These modifications affect the kinematic terms of the decomposition \eqref{jacord}, as well as the entries of the matrix \eqref{A15}.
Finally, the anomalous terms that enter the definition of the vector \eqref{phivec} are given, for $g^{\bullet}g^{\bullet}\to q\bar{q}g$, by

\newpage	

\begin{widetext}
\begin{align}
\phi_{[1,3,12]}&=\bar{u}_{3}\left[J_{\text{q}}\left(p_{1}+p_{2},p_{4},p_{3},p_{5}\right)_{\alpha\alpha_{5}}\right]v_{4}\Pi_{\text{Fey}}^{\alpha\beta}V_{\beta\alpha_{1}\alpha_{2}}\left(-p_{1}-p_{2},p_{1},p_{2}\right)\varepsilon_{1}^{\alpha_{1}}\varepsilon_{2}^{\alpha_{2}}\varepsilon_{5}^{\alpha_{5}}+n_{3;16}=s_{12}\varphi_{[1,3,12]},\nn
\phi_{[1,4,15]}&=\bar{u}_{3}\left[ J_{\text{g}^{\bullet}\text{g}^{\bullet}\bar{\text{q}}\text{q}}\left(p_{2},p_{4}+p_{5},p_{3},p_{1}\right)_{\alpha_{2}\alpha_{1}}\left(\slashed p_{4}+\slashed p_{5}\right)\slashed\varepsilon_{5}\right]v_{4}\,\varepsilon_{1}^{\alpha_{1}}\varepsilon_{2}^{\alpha_{2}}=s_{45}\varphi_{[1,4,15]},\nn
\phi_{[2,4,7]}&=\bar{u}_{3}\left[\slashed\varepsilon_{2}\left(\slashed p_{3}+\slashed p_{2}+i\mu\gamma^{5}\right)J_{\text{g}^{\bullet}\text{g}\bar{\text{q}}^{\bullet}\text{q}}\left(p_{1},p_{4},p_{3}+p_{2},p_{5}\right)_{\alpha_{1}\alpha_{5}}\right]v_{4}\,\varepsilon_{1}^{\alpha_{1}}\varepsilon_{5}^{\alpha_{5}}=s_{23}\varphi_{[2,4,7]},\nn
\phi_{[2,5,11]}&=\bar{u}_{3}\left[J_{\text{g}^{\bullet}\text{g}^{\bullet}\bar{\text{q}}\text{q}}\left(p_{1}+p_{5},p_{4},p_{3},p_{2}\right)_{\alpha\alpha_{2}}\right]v_{4}\,\Pi_{\text{gen}}^{\alpha\beta}\left(p_{1}+p_{5},\mu^{2}\right)V_{\beta\alpha_{1}\alpha_{5}}\left(-p_{1}-p_{5},p_{1},p_{5}\right)\varepsilon_{1}^{\alpha_{1}}\varepsilon_{2}^{\alpha_{2}}\varepsilon_{5}^{\alpha_{5}}+n_{5;16}=s_{15}\varphi_{[2,5,11]},\nn
\phi_{[6,7,14]}&=\bar{u}_{3}\left[J_{\text{g}^{\bullet}\text{g}\bar{\text{q}}^{\bullet}\text{q}}\left(p_{5},p_{4}+p_{1},p_{3},p_{2}\right)_{\alpha_{5}\alpha_{2}}\left(\slashed p_{4}+\slashed p_{1}+i\mu\gamma^{5}\right)\slashed\varepsilon_{1}\right]v_{4}\,\varepsilon_{2}^{\alpha_{2}}\varepsilon_{5}^{\alpha_{5}}=s_{14}\varphi_{[6,7,14]},\nn
\phi_{[6,8,9]}&=\bar{u}_{3}\left[J_{\text{g}^{\bullet}\text{g}^{\bullet}\bar{\text{q}}\text{q}}\left(p_{2}+p_{5},p_{4},p_{3},p_{1}\right)_{\alpha\alpha_{1}}\right]v_{4}\,\Pi_{\text{gen}}^{\alpha\beta}\left(p_{2}+p_{5},\mu^{2}\right)V_{\beta\alpha_{2}\alpha_{5}}\left(-p_{2}-p_{5},p_{2},p_{5}\right)\varepsilon_{1}^{\alpha_{1}}\varepsilon_{2}^{\alpha_{2}}\varepsilon_{5}^{\alpha_{5}}+n_{8;16}=s_{25}\varphi_{[6,8,9]},\nn
\phi_{[9,10,15]}&=\bar{u}_{3}\left[\slashed\varepsilon_{1}\left(\slashed p_{3}+\slashed p_{1}+i\mu\gamma^{5}\right)J_{\text{g}^{\bullet}\text{g}\bar{\text{q}}^{\bullet}\text{q}}\left(p_{5},p_{4},p_{3}+p_{1},p_{2}\right)_{\alpha_{5}\alpha_{2}}\right]v_{4}\,\varepsilon_{2}^{\alpha_{2}}\varepsilon_{5}^{\alpha_{5}}=s_{13}\varphi_{[9,10,15]},\nn
\phi_{[11,10,13]}&=\bar{u}_{3}\left[J_{\text{g}^{\bullet}\text{g}\bar{\text{q}}^{\bullet}\text{q}}\left(p_{1},p_{4}+p_{2},p_{3},p_{5}\right)_{\alpha_{1}\alpha_{5}}\left(\slashed p_{4}+\slashed p_{2}+i\mu\gamma^{5}\right)\slashed\varepsilon_{2}\right]v_{4}\,\varepsilon_{1}^{\alpha_{1}}\varepsilon_{5}^{\alpha_{5}}=s_{24}\varphi_{[11,10,13]},\nn
\phi_{[12,13,14]}&=\bar{u}_{3}\left[\slashed\varepsilon_{5}\left(\slashed p_{3}+\slashed p_{5}\right)J_{\text{g}^{\bullet}\text{g}^{\bullet}\bar{\text{q}}\text{q}}\left(p_{2},p_{4},p_{3}+p_{5},p_{1}\right)_{\alpha_{2}\alpha_{1}}\right]v_{4}\,\varepsilon_{1}^{\alpha_{1}}\varepsilon_{2}^{\alpha_{2}}=s_{35}\varphi_{[12,13,14]},
\end{align}
 and, for $s^{ \bullet}s^{\bullet}\to q\bar{q}g$,
\begin{align}
\phi_{[1,3,12]}&=\bar{u}_{3}\left[J_{\text{q}}\left(p_{1}+p_{2},p_{4},p_{3},p_{5}\right)_{\alpha\alpha_{5}}\right]v_{4}\Pi_{\text{Fey}}^{\alpha\beta}\,\left(p_{1}-p_{2}\right)_{\beta}\varepsilon_{5}^{\alpha_{5}}=s_{12}\varphi_{[1,3,12]},\nn
\phi_{[1,4,15]}&=\bar{u}_{3}\left[J_{\text{s}^{\bullet}\text{s}^{\bullet}\bar{\text{q}}\text{q}}\left(p_{2},p_{4}+p_{5},p_{3},p_{1}\right)\left(\slashed p_{4}+\slashed p_{5}\right)\slashed\varepsilon_{5}\right]v_{4}=s_{45}\varphi_{[1,4,15]},\nn
\phi_{[2,4,7]}&=\bar{u}_{3}\left[\gamma^{5}\left(\slashed p_{3}+\slashed p_{2}+i\mu\gamma^{5}\right)J_{\text{s}^{\bullet}\text{g}\bar{\text{q}}^{\bullet}\text{q}}\left(p_{1},p_{4},p_{3}+p_{2},p_{5}\right)_{\alpha_{5}}\right]v_{4}\,\varepsilon_{5}^{\alpha_{5}}=s_{23}\varphi_{[2,4,7]},\nn
\phi_{[2,5,11]}&=\bar{u}_{3}\left[J_{\text{s}^{\bullet}\text{s}^{\bullet}\bar{\text{q}}\text{q}}\left(p_{1}+p_{5},p_{4},p_{3},p_{2}\right)\right]v_{4}\left(2p_{1}+p_{5}\right)\cdot\varepsilon_{5}+n_{5;16}=s_{15}\varphi_{[2,5,11]},\nn
\phi_{[6,7,14]}&=\bar{u}_{3}\left[J_{\text{s}^{\bullet}\text{g}\bar{\text{q}}^{\bullet}\text{q}}\left(p_{5},p_{4}+p_{1},p_{3},p_{2}\right)_{\alpha_{5}}\left(\slashed p_{4}+\slashed p_{1}+i\mu\gamma^{5}\right)\slashed\varepsilon_{1}\right]v_{4}\,\varepsilon_{5}^{\alpha_{5}}=s_{14}\varphi_{[6,7,14]},\nn
\phi_{[6,8,9]}&=\bar{u}_{3}\left[J_{\text{s}^{\bullet}\text{s}^{\bullet}\bar{\text{q}}\text{q}}\left(p_{2}+p_{5},p_{4},p_{3},p_{1}\right)\right]v_{4}\left(2p_{2}+p_{5}\right)\cdot\varepsilon_{5}+n_{8;16}=s_{25}\varphi_{[6,8,9]},\nn
\phi_{[9,10,15]}&=\bar{u}_{3}\left[\gamma^{5}\left(\slashed p_{3}+\slashed p_{1}+i\mu\gamma^{5}\right)J_{\text{s}^{\bullet}\text{g}\bar{\text{q}}^{\bullet}\text{q}}\left(p_{5},p_{4},p_{3}+p_{1},p_{2}\right)_{\alpha_{5}}\right]v_{4}\,\varepsilon_{5}^{\alpha_{5}}=s_{13}\varphi_{[9,10,15]},\nn
\phi_{[11,10,13]}&=\bar{u}_{3}\left[J_{\text{s}^{\bullet}\text{g}\bar{\text{q}}^{\bullet}\text{q}}\left(p_{1},p_{4}+p_{2},p_{3},p_{5}\right)_{\alpha_{5}}\left(\slashed p_{4}+\slashed p_{2}+i\mu\gamma^{5}\right)\gamma^{5}\right]v_{4}\,\varepsilon_{5}^{\alpha_{5}}=s_{24}\varphi_{[11,10,13]},\nn
\phi_{[12,13,14]}&=\bar{u}_{3}\left[\slashed\varepsilon_{5}\left(\slashed p_{3}+\slashed p_{5}\right)J_{\text{s}^{\bullet}\text{s}^{\bullet}\bar{\text{q}}\text{q}}\left(p_{2},p_{4},p_{3}+p_{5},p_{1}\right)\right]v_{4}=s_{35}\varphi_{[12,13,14]}.
\end{align}
\end{widetext}

\noindent The systems $\mathbb{A}\Delta=\boldsymbol\phi$ associated to both amplitudes still satisfy the condition \eqref{rank}, which ensures the existence of a C/K-dual representation, depending on four arbitrary parameters, whose expression follows the structure of \eqref{solution}.\\
 As for the pure four-dimensional case, the analytic expressions of the dual numerators  have been obtained for generic polarizations and numerical checks of the result have been performed for different helicity configurations, including longitudinal polarizations of generalized gluons.


\section{Conclusions}
In this letter we investigated, from a diagrammatic point of view, the
off-shell color-kinematics duality for amplitudes in gauge theories
coupled with matter in four as well as $d$ dimensions, within the Four
Dimensional Formulation variant of the Four Dimensional Helicity scheme. 
This duality, first observed at tree-level for on-shell four-point amplitudes, is non-trivially satisfied within higher-multiplicity tree-level or multi-loop graphs, due to presence of contact terms which violate the Jacobi identity for numerators. We studied the source of such anomalous terms in $gg \to ss, q {\bar q}, gg$ scattering processes.
Working in axial gauge, we have explicitly shown that, whenever the Lie structure constants obey a Jacobi identity, the analogous combination of their kinematic numerators can {\it always} be  reduced to a sum of 
numerators of sub-diagrams, with one or two denominators less. 

Our decomposition provides a systematic classification of the duality-violating terms into a reduced number of effective vertices and, since they vanish when on-shellness is imposed on the four particles identifying the Jacobi relation, it immediately allows to recover the color-kinematics duality for multi-loop cut-integrands built from Feynman rules. 

We consider this study as an independent step towards a different, yet direct perspective to the diagrammatic understanding of color-dual graphs. Our approach, based on the direct inspection of Feynman diagrams and the identification of a set of constraints able to remove C/K-violating terms,
offers a method for the construction of dual numerators which is alternative 
to the traditional one, where, starting from general ansatz on the functional dependence of numerators on external momenta and polarizations,
Jacobi-like symmetries are imposed.

Off-shell recurrence relations, in tandem with generalized gauge transformations, can play an important role 
for gaining further understanding of the on-shell C/K-duality in
higher-multiplicity processes, as shown by our explicit determination
of dual numerators for the tree-level $gg\to q\bar{q}g$ amplitude,
first computed in four dimensions and later in $d$ dimensions, where
the initial state gluons were considered as dimensionally regulated
particles.

We expect this approach to have a natural extension at loop-level, which will be object of future studies.


\section*{Acknowledgements}
We wish to thank John Joseph Carrasco and Hui Luo for clarifying
discussions and comments on the manuscript. We also acknowledge 
Lorenzo Bianchi, Scott Davies, Marco Di Mauro, Reinke Isermann, Adele Naddeo and Josh Nohle. W.J.T also acknowledges Giacomo Bighin for inspiring discussions regarding computing.\\
The work of P.M. and U.S. is supported by the Alexander von Humboldt
Foundation, in the framework of the Sofja Kovalevskaja Award 2010,
endowed by the German Federal Ministry of Education and Research.
W.J.T. is supported by Fondazione Cassa di Risparmio di Padova e Rovigo (CARIPARO).\\
This work is also partially supported by Padua University Project CPDA144437.\\
The Feynman diagrams depicted in this paper are generated using \Feynarts~\cite{Hahn:2000kx}.

\appendix
\section{}\label{AppA}
In this Appendix we briefly recall the main features of the FDF scheme and we give the set of Feynman rules used in the calculations discussed in Section \ref{FDFduality}.
\begin{itemize}
	\item We use barred notation for quantities referred to unobserved particles,  living in a $d$-dimensional space. Thus, the metric tensor 
	\begin{align}
		\bar g^{\mu \nu} = g^{\mu \nu} + \tilde g^{\mu \nu}  \, ,
	\end{align}
	can be decomposed in terms of a four-dimensional tensor  $g$ and a $-2 \epsilon$-dimensional one, $ \tilde g$.
	The  tensors $g$ and $\tilde g$  project a $d$-dimensional vector $\bar q$ into the four-dimensional  and the  
	$-2 \epsilon$-dimensional subspaces, respectively. 
	\item $d$-dimensional  momenta $\bar \ell$ are decomposed as
	\begin{align}
	\bar \ell = \ell + \tilde \ell \, , \qquad \bar \ell^2  = \ell^2 -\mu^2 = m^2  \,  
	\label{Eq:Dec0}
	\end{align}
	\item The algebra of matrices $\tilde \gamma^\mu = \tilde g^{\mu}_{\phantom{\mu} \nu} \, \bar \gamma^\nu$,
	\begin{subequations}
		\begin{align}
			[ \tilde \gamma^{\alpha}, \gamma^{5}   ] &= 0 \, , & 
			\{\tilde \gamma^{\alpha}, \gamma^{\mu}  \} &=0 \ , \label{Eq:Gamma01} \\
			\{\tilde \gamma^{\alpha}, \tilde \gamma^{\beta}  \} &= 2 \,  \tilde g^{\alpha \beta} \, . 
		\end{align}\label{Eq:Gamma02}
	\end{subequations}
	is implemented through the substitutions
	\begin{align}
		\tilde g^{\alpha \beta} \to   G^{AB}, \qquad  \tilde \ell^{\alpha} \to i \, \mu \, \QQ^A \; , \qquad  \tilde \gamma^\alpha \to \gamma^5 \, \GA^A\, .
		\label{Eq:SubF}
	\end{align} 
	together with the set of selection rules, ($-2\epsilon$)-SRs,
		\begin{align}
		\GG^{AB}\GG^{BC} &= \GG^{AC},    & \GG^{AA}&=0,  &   \GG^{AB}&=\GG^{BA},  \nn
		\GA^A \GG^{AB} &= \GA^B,              &   \GA^A \GA^{A} &=0,  & \QQ^A \GA^{A}  &=1, \nn
		\QQ^A \GG^{AB} &= \QQ^B,              & \QQ^A \QQ^{A} &=1.  
		\label{Eq:2epsA}
	\end{align}
	which, ensuring the exclusion of the terms containing odd powers of $\mu$, completely
	defines the FDF and allows the construction of integrands which, upon
	integration, yield to the same result as in the FDH scheme.
	\item
	The spinors of a $d$-dimensional fermion fulfil the completeness relations 
	\begin{align}
	\sum_{\lambda=\pm}u_{\lambda}\left(\ell \right)\bar{u}_{\lambda}\left(\ell \right) & = \slashed \ell + i \mu \gamma^5 + m \, , \nn
	\sum_{\lambda=\pm}v_{\lambda}\left(\ell  \right)\bar{v}_{\lambda}\left(\ell \right)  & = \slashed \ell + i \mu \gamma^5  - m \, ,
	\label{Eq:CompF4}
	\end{align}
	which consistently reconstruct the numerator of the cut  propagator. 
\item In the axial gauge, the helicity sum of a $d$-dimensional transverse polarization vector can be disentangled in
\begin{small}
	\begin{align}
	&\sum_{i=1}^{d -2} \, \varepsilon_{i\, (d)}^\mu\left (\bar \ell , \bar \eta \right )\varepsilon_{i\, (d)}^{\ast \nu}\left (\bar \ell , \bar \eta \right ) =\nn
	&\left (   - g^{\mu \nu}  +\frac{ \ell^\mu \ell^\nu}{\mu^2} \right) -\left (  \tilde g^{\mu \nu}  +
	\frac{ \tilde \ell^\mu \tilde \ell^\nu}{\mu^2} \right ) \, ,
	\label{Eq:CompGD2}
	\end{align}
\end{small}
\noindent where the first term can be regarded as the cut propagator of a massive vector boson,
\begin{align}
\sum_{\lambda=\pm,0}\varepsilon_{\lambda}^{\mu}(\ell) \, \varepsilon_{\lambda}^{*\nu}(\ell)&= -g^{\mu\nu}+\frac{\ell^{\mu}\ell^{\nu}}{\mu^{2}}  \, ,  \label{flat}
\end{align}
whose polarizations obey the expected properties
\begin{align}
&\varepsilon^2_{\pm}(\ell)  =\phantom{-} 0\, ,  & &\varepsilon_{\pm}(\ell)\cdot\varepsilon_{\mp}(\ell)=-1\, , \nn
&\varepsilon_{0}^2(\ell) =-1\, , &  &\varepsilon_{\pm}(\ell)\cdot\varepsilon_{0}(\ell)   =\phantom{-} 0\, ,  \nn
&\varepsilon_{\lambda}(\ell) \cdot \ell =\phantom{-} 0 \, .
\label{Eq:propEps}
\end{align}
For the explicit expression of polarization vectors as well as generalized spinor we refer the reader to~\cite{Fazio:2014xea}. 
The second term of the r.h.s. of Eq.~(\ref{Eq:CompGD2})  is related to the numerator of cut propagator of the scalar $s^{\bullet}$ and can be expressed in terms
of the $(-2 \epsilon)$-SRs as:
\begin{equation}
\tilde g^{\mu \nu}  +\frac{ \tilde \ell^\mu \tilde \ell^\nu}{\mu^2}  \quad \to  \quad    \GG^{AB} - \QQ^A \QQ^B  \, .
\label{Eq:Pref}
\end{equation}
\end{itemize}
Within FDF scheme, the QCD  $d$-dimensional Feynman rules in axial gauge have the following four-dimensional formulation:
\begin{subequations}
\vspace{-0.4cm}
\begin{align}
\parbox{20mm}{
\unitlength=0.20bp%
\begin{feynartspicture}(300,300)(1,1)
\FADiagram{}
\FAProp(4.,10.)(16.,10.)(0.,){/Cycles}{0}
\FALabel(5.5,8.93)[t]{\tiny $a, \alpha$}
\FALabel(14.5,8.93)[t]{\tiny $b, \beta$}
\FALabel(10.,12.5)[]{\tiny $k$}
\FAVert(4.,10.){0}
\FAVert(16.,10.){0}
\end{feynartspicture}} &= i \, \frac{ \delta^{ab}  }{k^2 -\mu^2 +i 0}\bigg(-g^{\alpha\beta}+\frac{k^{\alpha}k^{\beta}}{\mu^2}\bigg)\,,
\label{Eq:FRglu}\\[-3.0ex]
\parbox{20mm}{\unitlength=0.20bp%
\begin{feynartspicture}(300,300)(1,1)
\FADiagram{}
\FAProp(4.,10.)(16.,10.)(0.,){/ScalarDash}{0}
\FALabel(5.5,8.93)[t]{\tiny $a, A$}
\FALabel(14.5,8.93)[t]{\tiny $b, B$}
\FALabel(10.,12.5)[]{\tiny $k$}
\FAVert(4.,10.){0}
\FAVert(16.,10.){0}
\end{feynartspicture}} &= i \,  \,\frac{\delta^{ab}}{k^2 -\mu^2+ i0} \left(G^{AB}-Q^AQ^B\right)\, , 
\\[-3.0ex]
\parbox{20mm}{\unitlength=0.20bp%
\begin{feynartspicture}(300,300)(1,1)
\FADiagram{}
\FAProp(4.,10.)(16.,10.)(0.,){/Straight}{1}
\FALabel(5.5,8.93)[t]{\tiny $i$}
\FALabel(14.5,8.93)[t]{\tiny $j$}
\FALabel(10.,12.5)[]{\tiny $k$}
\FAVert(4.,10.){0}
\FAVert(16.,10.){0}
\end{feynartspicture}} &= i \, \delta^{ij} \,\frac{ \slashed k + i \mu \gamma^5 +m }{k^2 -m^2 -\mu^2+i0}  \, ,  
\label{Eq:FRfer}   \\[-3.0ex] 
 \parbox{20mm}{\unitlength=0.20bp%
\begin{feynartspicture}(300,300)(1,1)
\FADiagram{}
\FAProp(3.,10.)(10.,10.)(0.,){/Cycles}{0}
\FALabel(5.3,8.93)[t]{\tiny $1, a, \alpha$}
\FAProp(16.,15.)(10.,10.)(0.,){/Cycles}{0}
\FALabel(12.2273,13.5749)[br]{\tiny $2, b, \beta$}
\FAProp(16.,5.)(10.,10.)(0.,){/Cycles}{0}
\FALabel(12.8873,5.81315)[tr]{\tiny $3, c, \gamma$}
\FAVert(10.,10.){0}
\end{feynartspicture}} &=  -g \, f^{abc} \, \big [  (k_1-k_2)^\gamma g^{\alpha\beta}\nonumber  \\[-4.0ex]
&\qquad +(k_2-k_3)^\alpha g^{\beta\gamma} + (k_3-k_1)^\beta g^{\gamma\alpha} \big ]  \, ,   \\[-3.0ex]
 \parbox{20mm}{\unitlength=0.20bp%
\begin{feynartspicture}(300,300)(1,1)
\FADiagram{}
\FAProp(3.,10.)(10.,10.)(0.,){/Cycles}{0}
\FALabel(5.3,8.93)[t]{\tiny $1,a, \alpha$}
\FAProp(16.,15.)(10.,10.)(0.,){/ScalarDash}{0}
\FALabel(12.2273,13.5749)[br]{\tiny $2, b, B$}
\FAProp(16.,5.)(10.,10.)(0.,){/ScalarDash}{0}
\FALabel(12.8873,5.81315)[tr]{\tiny $3,c, C$}
\FAVert(10.,10.){0}
\end{feynartspicture}} &=  -g \, f^{abc} \, (k_2-k_3)^\alpha  \, \GG^{BC}  \, ,  \\[-3.0ex]
%
 \parbox{20mm}{\unitlength=0.20bp%
\begin{feynartspicture}(300,300)(1,1)
\FADiagram{}
\FAProp(3.,10.)(10.,10.)(0.,){/Cycles}{0}
\FALabel(5.3,8.93)[t]{\tiny $1, a, \alpha$}
\FAProp(16.,15.)(10.,10.)(0.,){/ScalarDash}{0}
\FALabel(12.2273,13.5749)[br]{\tiny $2, b, B$}
\FAProp(16.,5.)(10.,10.)(0.,){/Cycles}{0}
\FALabel(12.8873,5.81315)[tr]{\tiny $3, c, \gamma$}
\FAVert(10.,10.){0}
\end{feynartspicture}} &= \mp g \, f^{abc} \,   ( i \mu) \, g^{\gamma\alpha}\, \QQ^B \nonumber  \\[-4.0ex]
&\qquad\qquad\qquad    (\tilde k_1 = 0, \quad  \tilde k_3 =\pm \tilde \ell) ,  \label{Eq:FRggs} \\[-3.ex]
%
\parbox{20mm}{\unitlength=0.20bp%
\begin{feynartspicture}(300,300)(1,1)
\FADiagram{}
\FAProp(4.,15.)(10.,10.)(0.,){/Cycles}{0}
\FALabel(3.0,12.)[]{\tiny $1,a,\alpha$}
\FAProp(10.,10.)(4.,5.)(0.,){/Cycles}{0}
\FALabel(3.0,8.)[]{\tiny $4,d,\delta$}
\FAProp(16.,15.)(10.,10.)(0.,){/Cycles}{0}
\FALabel(11.,15.)[]{\tiny $2,b,\beta$}
\FAProp(10.,10.)(16.,5.)(0.,){/Cycles}{0}
\FALabel(11,5.)[]{\tiny $3,c,\gamma$}
\FAVert(10.,10.){0}
\end{feynartspicture}} &= - i g^2 \Big [       f^{xad} \, f^{xbc} \, \left ( g^{\alpha \beta} g^{\delta \gamma} - g^{\alpha\gamma} g^{\beta \delta}\right )     \nonumber  \\[-4.0ex]
   &\qquad +        f^{xac} \, f^{xbd} \, \left ( g^{\alpha \beta} g^{\delta \gamma} - g^{\alpha\delta} g^{\beta \gamma}\right )  \nn 
   &\qquad +        f^{xab} \, f^{xdc} \, \left ( g^{\alpha \delta} g^{\beta \gamma} - g^{\alpha\gamma} g^{\beta \delta}\right )  \Big ]  \, ,  \\[-3.0ex]
\parbox{20mm}{\unitlength=0.20bp%
\begin{feynartspicture}(300,300)(1,1)
\FADiagram{}
\FAProp(4.,15.)(10.,10.)(0.,){/Cycles}{0}
\FALabel(3.0,12.)[]{\tiny $1,a,\alpha$}
\FAProp(10.,10.)(4.,5.)(0.,){/Cycles}{0}
\FALabel(3.0,8.)[]{\tiny $4,d,\delta$}
\FAProp(16.,15.)(10.,10.)(0.,){/ScalarDash}{0}
\FALabel(11.,15.)[]{\tiny $2,b,B$}
\FAProp(10.,10.)(16.,5.)(0.,){/ScalarDash}{0}
\FALabel(11,5.)[]{\tiny $3,c,C$}
\FAVert(10.,10.){0}
\end{feynartspicture}} &=  i g^2 \, g^{\alpha \delta} \; \big (  f^{xab} \, f^{xcd}  + f^{xac} \, f^{xbd} \big ) \; \GG^{BC}  \, ,  \\[-3.ex]
%
 \parbox{20mm}{\unitlength=0.20bp%
\begin{feynartspicture}(300,300)(1,1)
\FADiagram{}
\FAProp(3.,10.)(10.,10.)(0.,){/Straight}{1}
\FALabel(5.3,8.93)[t]{\tiny $1, i$}
\FAProp(16.,15.)(10.,10.)(0.,){/Cycles}{0}
\FALabel(12.2273,13.5749)[br]{\tiny $2, b, \beta$}
\FAProp(16.,5.)(10.,10.)(0.,){/Straight}{-1}
\FALabel(12.8873,5.81315)[tr]{\tiny $3, j$}
\FAVert(10.,10.){0}
\end{feynartspicture}} &= - i g \, \left ( t^{b}\right)_{ji} \,\gamma^\beta \, ,  \\[-3.0ex]
\parbox{20mm}{\unitlength=0.20bp%
\begin{feynartspicture}(300,300)(1,1)
\FADiagram{}
\FAProp(3.,10.)(10.,10.)(0.,){/Straight}{1}
\FALabel(5.3,8.93)[t]{\tiny $1, i$}
\FAProp(16.,15.)(10.,10.)(0.,){/ScalarDash}{0}
\FALabel(12.2273,13.5749)[br]{\tiny $2, b, B$}
\FAProp(16.,5.)(10.,10.)(0.,){/Straight}{-1}
\FALabel(12.8873,5.81315)[tr]{\tiny $3, j$}
\FAVert(10.,10.){0}
\end{feynartspicture}} &= - i g \, \left ( t^{b}\right)_{ji} \,\gamma^5 \,  \GA^B  \, .
\end{align}
\label{Eq:FR4}
\end{subequations}
In the Feynman rules~(\ref{Eq:FR4})  all the momenta are incoming and   the scalar particle  $s^{\bullet}$ can circulate in  the loop only. 
The $\mu^2$ term appearing in the propagators~(\ref{Eq:FRglu})--(\ref{Eq:FRfer}) enter only if the corresponding momentum 
$k$ is $d$-dimensional,  {\it i.e.} only if  the corresponding particle  circulates in  the loop.  In the vertex~(\ref{Eq:FRggs}) the momentum 
$k_1$ is four-dimensional,  while the other two are $d$-dimensional.


\bibliographystyle{elsarticle-num}
\bibliography{references}

\begin{thebibliography}{10}
\expandafter\ifx\csname url\endcsname\relax
  \def\url#1{\texttt{#1}}\fi
\expandafter\ifx\csname urlprefix\endcsname\relax\def\urlprefix{URL }\fi
\expandafter\ifx\csname href\endcsname\relax
  \def\href#1#2{#2} \def\path#1{#1}\fi

\bibitem{Bern:2008qj}
Z.~Bern, J.~Carrasco, H.~Johansson, {New Relations for Gauge-Theory
  Amplitudes}, Phys.Rev. D78 (2008) 085011.
\newblock \href {http://arxiv.org/abs/0805.3993} {\path{arXiv:0805.3993}},
  \href {http://dx.doi.org/10.1103/PhysRevD.78.085011}
  {\path{doi:10.1103/PhysRevD.78.085011}}.

\bibitem{Bern:2010ue}
Z.~Bern, J.~J.~M. Carrasco, H.~Johansson, {Perturbative Quantum Gravity as a
  Double Copy of Gauge Theory}, Phys.Rev.Lett. 105 (2010) 061602.
\newblock \href {http://arxiv.org/abs/1004.0476} {\path{arXiv:1004.0476}},
  \href {http://dx.doi.org/10.1103/PhysRevLett.105.061602}
  {\path{doi:10.1103/PhysRevLett.105.061602}}.

\bibitem{Johansson:2014zca}
H.~Johansson, A.~Ochirov, {Pure Gravities via Color-Kinematics Duality for
  Fundamental Matter}\href {http://arxiv.org/abs/1407.4772}
  {\path{arXiv:1407.4772}}.

\bibitem{Naculich:2014naa}
S.~G. Naculich, {Scattering equations and BCJ relations for gauge and
  gravitational amplitudes with massive scalar particles}, JHEP 09 (2014) 029.
\newblock \href {http://arxiv.org/abs/1407.7836} {\path{arXiv:1407.7836}},
  \href {http://dx.doi.org/10.1007/JHEP09(2014)029}
  {\path{doi:10.1007/JHEP09(2014)029}}.

\bibitem{Johansson:2015oia}
H.~Johansson, A.~Ochirov, {Color-Kinematics Duality for QCD Amplitudes}\href
  {http://arxiv.org/abs/1507.00332} {\path{arXiv:1507.00332}}.

\bibitem{Bern:2011ia}
Z.~Bern, T.~Dennen, {A Color Dual Form for Gauge-Theory Amplitudes},
  Phys.Rev.Lett. 107 (2011) 081601.
\newblock \href {http://arxiv.org/abs/1103.0312} {\path{arXiv:1103.0312}},
  \href {http://dx.doi.org/10.1103/PhysRevLett.107.081601}
  {\path{doi:10.1103/PhysRevLett.107.081601}}.

\bibitem{Bern:2010yg}
Z.~Bern, T.~Dennen, Y.-t. Huang, M.~Kiermaier, {Gravity as the Square of Gauge
  Theory}, Phys.Rev. D82 (2010) 065003.
\newblock \href {http://arxiv.org/abs/1004.0693} {\path{arXiv:1004.0693}},
  \href {http://dx.doi.org/10.1103/PhysRevD.82.065003}
  {\path{doi:10.1103/PhysRevD.82.065003}}.

\bibitem{Bern:2013yya}
Z.~Bern, S.~Davies, T.~Dennen, Y.-t. Huang, J.~Nohle, {Color-Kinematics Duality
  for Pure Yang-Mills and Gravity at One and Two Loops}\href
  {http://arxiv.org/abs/1303.6605} {\path{arXiv:1303.6605}}.

\bibitem{BjerrumBohr:2009rd}
N.~Bjerrum-Bohr, P.~H. Damgaard, P.~Vanhove, {Minimal Basis for Gauge Theory
  Amplitudes}, Phys.Rev.Lett. 103 (2009) 161602.
\newblock \href {http://arxiv.org/abs/0907.1425} {\path{arXiv:0907.1425}},
  \href {http://dx.doi.org/10.1103/PhysRevLett.103.161602}
  {\path{doi:10.1103/PhysRevLett.103.161602}}.

\bibitem{BjerrumBohr:2010zs}
N.~Bjerrum-Bohr, P.~H. Damgaard, T.~Sondergaard, P.~Vanhove, {Monodromy and
  Jacobi-like Relations for Color-Ordered Amplitudes}, JHEP 1006 (2010) 003.
\newblock \href {http://arxiv.org/abs/1003.2403} {\path{arXiv:1003.2403}},
  \href {http://dx.doi.org/10.1007/JHEP06(2010)003}
  {\path{doi:10.1007/JHEP06(2010)003}}.

\bibitem{BjerrumBohr:2010yc}
N.~Bjerrum-Bohr, P.~H. Damgaard, B.~Feng, T.~Sondergaard, {Proof of Gravity and
  Yang-Mills Amplitude Relations}, JHEP 1009 (2010) 067.
\newblock \href {http://arxiv.org/abs/1007.3111} {\path{arXiv:1007.3111}},
  \href {http://dx.doi.org/10.1007/JHEP09(2010)067}
  {\path{doi:10.1007/JHEP09(2010)067}}.

\bibitem{BjerrumBohr:2010hn}
N.~Bjerrum-Bohr, P.~H. Damgaard, T.~Sondergaard, P.~Vanhove, {The Momentum
  Kernel of Gauge and Gravity Theories}, JHEP 1101 (2011) 001.
\newblock \href {http://arxiv.org/abs/1010.3933} {\path{arXiv:1010.3933}},
  \href {http://dx.doi.org/10.1007/JHEP01(2011)001}
  {\path{doi:10.1007/JHEP01(2011)001}}.

\bibitem{BjerrumBohr:2012mg}
N.~Bjerrum-Bohr, P.~H. Damgaard, R.~Monteiro, D.~O'Connell, {Algebras for
  Amplitudes}, JHEP 1206 (2012) 061.
\newblock \href {http://arxiv.org/abs/1203.0944} {\path{arXiv:1203.0944}},
  \href {http://dx.doi.org/10.1007/JHEP06(2012)061}
  {\path{doi:10.1007/JHEP06(2012)061}}.

\bibitem{Stieberger:2009hq}
S.~Stieberger, {Open \& Closed vs. Pure Open String Disk Amplitudes}\href
  {http://arxiv.org/abs/0907.2211} {\path{arXiv:0907.2211}}.

\bibitem{Mafra:2010jq}
C.~R. Mafra, O.~Schlotterer, S.~Stieberger, D.~Tsimpis, {A recursive method for
  SYM n-point tree amplitudes}, Phys.Rev. D83 (2011) 126012.
\newblock \href {http://arxiv.org/abs/1012.3981} {\path{arXiv:1012.3981}},
  \href {http://dx.doi.org/10.1103/PhysRevD.83.126012}
  {\path{doi:10.1103/PhysRevD.83.126012}}.

\bibitem{Mafra:2011kj}
C.~R. Mafra, O.~Schlotterer, S.~Stieberger, {Explicit BCJ Numerators from Pure
  Spinors}, JHEP 1107 (2011) 092.
\newblock \href {http://arxiv.org/abs/1104.5224} {\path{arXiv:1104.5224}},
  \href {http://dx.doi.org/10.1007/JHEP07(2011)092}
  {\path{doi:10.1007/JHEP07(2011)092}}.

\bibitem{Mafra:2011nv}
C.~R. Mafra, O.~Schlotterer, S.~Stieberger, {Complete N-Point Superstring Disk
  Amplitude I. Pure Spinor Computation}, Nucl.Phys. B873 (2013) 419--460.
\newblock \href {http://arxiv.org/abs/1106.2645} {\path{arXiv:1106.2645}},
  \href {http://dx.doi.org/10.1016/j.nuclphysb.2013.04.023}
  {\path{doi:10.1016/j.nuclphysb.2013.04.023}}.

\bibitem{Grassi:2011ky}
P.~Grassi, A.~Mezzalira, L.~Sommovigo, {BCJ and KK Relations from BRST Symmetry
  and Supergravity Amplitudes}\href {http://arxiv.org/abs/1111.0544}
  {\path{arXiv:1111.0544}}.

\bibitem{Sondergaard:2011iv}
T.~Sondergaard, {Perturbative Gravity and Gauge Theory Relations: A Review},
  Adv.High Energy Phys. 2012 (2012) 726030.
\newblock \href {http://arxiv.org/abs/1106.0033} {\path{arXiv:1106.0033}},
  \href {http://dx.doi.org/10.1155/2012/726030}
  {\path{doi:10.1155/2012/726030}}.

\bibitem{Feng:2010my}
B.~Feng, R.~Huang, Y.~Jia, {Gauge Amplitude Identities by On-shell Recursion
  Relation in S-matrix Program}, Phys.Lett. B695 (2011) 350--353.
\newblock \href {http://arxiv.org/abs/1004.3417} {\path{arXiv:1004.3417}},
  \href {http://dx.doi.org/10.1016/j.physletb.2010.11.011}
  {\path{doi:10.1016/j.physletb.2010.11.011}}.

\bibitem{Jia:2010nz}
Y.~Jia, R.~Huang, C.-Y. Liu, {$U(1)$-decoupling, KK and BCJ relations in
  $\mathcal{N}=4$ SYM}, Phys.Rev. D82 (2010) 065001.
\newblock \href {http://arxiv.org/abs/1005.1821} {\path{arXiv:1005.1821}},
  \href {http://dx.doi.org/10.1103/PhysRevD.82.065001}
  {\path{doi:10.1103/PhysRevD.82.065001}}.

\bibitem{Chen:2011jxa}
Y.-X. Chen, Y.-J. Du, B.~Feng, {A Proof of the Explicit Minimal-basis Expansion
  of Tree Amplitudes in Gauge Field Theory}, JHEP 1102 (2011) 112.
\newblock \href {http://arxiv.org/abs/1101.0009} {\path{arXiv:1101.0009}},
  \href {http://dx.doi.org/10.1007/JHEP02(2011)112}
  {\path{doi:10.1007/JHEP02(2011)112}}.

\bibitem{Du:2011js}
Y.-J. Du, B.~Feng, C.-H. Fu, {BCJ Relation of Color Scalar Theory and KLT
  Relation of Gauge Theory}, JHEP 1108 (2011) 129.
\newblock \href {http://arxiv.org/abs/1105.3503} {\path{arXiv:1105.3503}},
  \href {http://dx.doi.org/10.1007/JHEP08(2011)129}
  {\path{doi:10.1007/JHEP08(2011)129}}.

\bibitem{Du:2011se}
Y.-J. Du, B.~Feng, C.-H. Fu, {Note on Permutation Sum of Color-ordered Gluon
  Amplitudes}, Phys.Lett. B706 (2012) 490--494.
\newblock \href {http://arxiv.org/abs/1110.4683} {\path{arXiv:1110.4683}},
  \href {http://dx.doi.org/10.1016/j.physletb.2011.11.063}
  {\path{doi:10.1016/j.physletb.2011.11.063}}.

\bibitem{Fu:2012uy}
C.-H. Fu, Y.-J. Du, B.~Feng, {An algebraic approach to BCJ numerators}, JHEP
  1303 (2013) 050.
\newblock \href {http://arxiv.org/abs/1212.6168} {\path{arXiv:1212.6168}},
  \href {http://dx.doi.org/10.1007/JHEP03(2013)050}
  {\path{doi:10.1007/JHEP03(2013)050}}.

\bibitem{Du:2013sha}
Y.-J. Du, B.~Feng, C.-H. Fu, {The Construction of Dual-trace Factor in
  Yang-Mills Theory}, JHEP 1307 (2013) 057.
\newblock \href {http://arxiv.org/abs/1304.2978} {\path{arXiv:1304.2978}},
  \href {http://dx.doi.org/10.1007/JHEP07(2013)057}
  {\path{doi:10.1007/JHEP07(2013)057}}.

\bibitem{Vaman:2010ez}
D.~Vaman, Y.-P. Yao, {Constraints and Generalized Gauge Transformations on
  Tree-Level Gluon and Graviton Amplitudes}, JHEP 1011 (2010) 028.
\newblock \href {http://arxiv.org/abs/1007.3475} {\path{arXiv:1007.3475}},
  \href {http://dx.doi.org/10.1007/JHEP11(2010)028}
  {\path{doi:10.1007/JHEP11(2010)028}}.

\bibitem{Boels:2011mn}
R.~H. Boels, R.~S. Isermann, {Yang-Mills amplitude relations at loop level from
  non-adjacent BCFW shifts}, JHEP 1203 (2012) 051.
\newblock \href {http://arxiv.org/abs/1110.4462} {\path{arXiv:1110.4462}},
  \href {http://dx.doi.org/10.1007/JHEP03(2012)051}
  {\path{doi:10.1007/JHEP03(2012)051}}.

\bibitem{Boels:2012ew}
R.~H. Boels, B.~A. Kniehl, O.~V. Tarasov, G.~Yang, {Color-kinematic Duality for
  Form Factors}, JHEP 1302 (2013) 063.
\newblock \href {http://arxiv.org/abs/1211.7028} {\path{arXiv:1211.7028}},
  \href {http://dx.doi.org/10.1007/JHEP02(2013)063}
  {\path{doi:10.1007/JHEP02(2013)063}}.

\bibitem{Boels:2013bi}
R.~H. Boels, R.~S. Isermann, R.~Monteiro, D.~O'Connell, {Colour-Kinematics
  Duality for One-Loop Rational Amplitudes}, JHEP 1304 (2013) 107.
\newblock \href {http://arxiv.org/abs/1301.4165} {\path{arXiv:1301.4165}},
  \href {http://dx.doi.org/10.1007/JHEP04(2013)107}
  {\path{doi:10.1007/JHEP04(2013)107}}.

\bibitem{Oxburgh:2012zr}
S.~Oxburgh, C.~White, {BCJ duality and the double copy in the soft limit}, JHEP
  1302 (2013) 127.
\newblock \href {http://arxiv.org/abs/1210.1110} {\path{arXiv:1210.1110}},
  \href {http://dx.doi.org/10.1007/JHEP02(2013)127}
  {\path{doi:10.1007/JHEP02(2013)127}}.

\bibitem{Saotome:2012vy}
R.~Saotome, R.~Akhoury, {Relationship Between Gravity and Gauge Scattering in
  the High Energy Limit}, JHEP 1301 (2013) 123.
\newblock \href {http://arxiv.org/abs/1210.8111} {\path{arXiv:1210.8111}},
  \href {http://dx.doi.org/10.1007/JHEP01(2013)123}
  {\path{doi:10.1007/JHEP01(2013)123}}.

\bibitem{Broedel:2011pd}
J.~Broedel, J.~J.~M. Carrasco, {Virtuous Trees at Five and Six Points for
  Yang-Mills and Gravity}, Phys.Rev. D84 (2011) 085009.
\newblock \href {http://arxiv.org/abs/1107.4802} {\path{arXiv:1107.4802}},
  \href {http://dx.doi.org/10.1103/PhysRevD.84.085009}
  {\path{doi:10.1103/PhysRevD.84.085009}}.

\bibitem{Broedel:2012rc}
J.~Broedel, L.~J. Dixon, {Color-kinematics duality and double-copy construction
  for amplitudes from higher-dimension operators}, JHEP 1210 (2012) 091.
\newblock \href {http://arxiv.org/abs/1208.0876} {\path{arXiv:1208.0876}},
  \href {http://dx.doi.org/10.1007/JHEP10(2012)091}
  {\path{doi:10.1007/JHEP10(2012)091}}.

\bibitem{Cachazo:2012uq}
F.~Cachazo, {Fundamental BCJ Relation in N=4 SYM From The Connected
  Formulation}\href {http://arxiv.org/abs/1206.5970} {\path{arXiv:1206.5970}}.

\bibitem{Monteiro:2011pc}
R.~Monteiro, D.~O'Connell, {The Kinematic Algebra From the Self-Dual Sector},
  JHEP 1107 (2011) 007.
\newblock \href {http://arxiv.org/abs/1105.2565} {\path{arXiv:1105.2565}},
  \href {http://dx.doi.org/10.1007/JHEP07(2011)007}
  {\path{doi:10.1007/JHEP07(2011)007}}.

\bibitem{Boels:2012sy}
R.~H. Boels, R.~S. Isermann, {On powercounting in perturbative quantum gravity
  theories through color-kinematic duality}, JHEP 1306 (2013) 017.
\newblock \href {http://arxiv.org/abs/1212.3473} {\path{arXiv:1212.3473}},
  \href {http://dx.doi.org/10.1007/JHEP06(2013)017}
  {\path{doi:10.1007/JHEP06(2013)017}}.

\bibitem{Carrasco:2015iwa}
J.~J.~M. Carrasco, {Gauge and Gravity Amplitude Relations}\href
  {http://arxiv.org/abs/1506.00974} {\path{arXiv:1506.00974}}.

\bibitem{Tolotti:2013caa}
M.~Tolotti, S.~Weinzierl, {Construction of an effective Yang-Mills Lagrangian
  with manifest BCJ duality}, JHEP 1307 (2013) 111.
\newblock \href {http://arxiv.org/abs/1306.2975} {\path{arXiv:1306.2975}},
  \href {http://dx.doi.org/10.1007/JHEP07(2013)111}
  {\path{doi:10.1007/JHEP07(2013)111}}.

\bibitem{Du:2012mt}
Y.-J. Du, H.~Luo, {On General BCJ Relation at One-loop Level in Yang-Mills
  Theory}, JHEP 1301 (2013) 129.
\newblock \href {http://arxiv.org/abs/1207.4549} {\path{arXiv:1207.4549}},
  \href {http://dx.doi.org/10.1007/JHEP01(2013)129}
  {\path{doi:10.1007/JHEP01(2013)129}}.

\bibitem{Nohle:2013bfa}
J.~Nohle, {Color-Kinematics Duality in One-Loop Four-Gluon Amplitudes with
  Matter}, Phys.Rev. D90~(2) (2014) 025020.
\newblock \href {http://arxiv.org/abs/1309.7416} {\path{arXiv:1309.7416}},
  \href {http://dx.doi.org/10.1103/PhysRevD.90.025020}
  {\path{doi:10.1103/PhysRevD.90.025020}}.

\bibitem{Bern:2012uf}
Z.~Bern, J.~Carrasco, L.~Dixon, H.~Johansson, R.~Roiban, {Simplifying Multiloop
  Integrands and Ultraviolet Divergences of Gauge Theory and Gravity
  Amplitudes}, Phys.Rev. D85 (2012) 105014.
\newblock \href {http://arxiv.org/abs/1201.5366} {\path{arXiv:1201.5366}},
  \href {http://dx.doi.org/10.1103/PhysRevD.85.105014}
  {\path{doi:10.1103/PhysRevD.85.105014}}.

\bibitem{Carrasco:2011mn}
J.~J. Carrasco, H.~Johansson, {Five-Point Amplitudes in N=4 Super-Yang-Mills
  Theory and N=8 Supergravity}, Phys.Rev. D85 (2012) 025006.
\newblock \href {http://arxiv.org/abs/1106.4711} {\path{arXiv:1106.4711}},
  \href {http://dx.doi.org/10.1103/PhysRevD.85.025006}
  {\path{doi:10.1103/PhysRevD.85.025006}}.

\bibitem{Bjerrum-Bohr:2013iza}
N.~E.~J. Bjerrum-Bohr, T.~Dennen, R.~Monteiro, D.~O'Connell, {Integrand
  Oxidation and One-Loop Colour-Dual Numerators in N=4 Gauge Theory}, JHEP 1307
  (2013) 092.
\newblock \href {http://arxiv.org/abs/1303.2913} {\path{arXiv:1303.2913}},
  \href {http://dx.doi.org/10.1007/JHEP07(2013)092}
  {\path{doi:10.1007/JHEP07(2013)092}}.

\bibitem{Carrasco:2012ca}
J.~J.~M. Carrasco, M.~Chiodaroli, M.~G{\"u}naydin, R.~Roiban, {One-loop
  four-point amplitudes in pure and matter-coupled N <= 4 supergravity}, JHEP
  1303 (2013) 056.
\newblock \href {http://arxiv.org/abs/1212.1146} {\path{arXiv:1212.1146}},
  \href {http://dx.doi.org/10.1007/JHEP03(2013)056}
  {\path{doi:10.1007/JHEP03(2013)056}}.

\bibitem{Bern:2011rj}
Z.~Bern, C.~Boucher-Veronneau, H.~Johansson, {N >= 4 Supergravity Amplitudes
  from Gauge Theory at One Loop}, Phys.Rev. D84 (2011) 105035.
\newblock \href {http://arxiv.org/abs/1107.1935} {\path{arXiv:1107.1935}},
  \href {http://dx.doi.org/10.1103/PhysRevD.84.105035}
  {\path{doi:10.1103/PhysRevD.84.105035}}.

\bibitem{BoucherVeronneau:2011qv}
C.~Boucher-Veronneau, L.~Dixon, {N >- 4 Supergravity Amplitudes from Gauge
  Theory at Two Loops}, JHEP 1112 (2011) 046.
\newblock \href {http://arxiv.org/abs/1110.1132} {\path{arXiv:1110.1132}},
  \href {http://dx.doi.org/10.1007/JHEP12(2011)046}
  {\path{doi:10.1007/JHEP12(2011)046}}.

\bibitem{Mastrolia:2012wf}
P.~Mastrolia, E.~Mirabella, G.~Ossola, T.~Peraro, {Integrand-Reduction for
  Two-Loop Scattering Amplitudes through Multivariate Polynomial Division},
  Phys. Rev. D87~(8) (2013) 085026.
\newblock \href {http://arxiv.org/abs/1209.4319} {\path{arXiv:1209.4319}},
  \href {http://dx.doi.org/10.1103/PhysRevD.87.085026}
  {\path{doi:10.1103/PhysRevD.87.085026}}.

\bibitem{Schubert:2014paa}
U.~Schubert, {Scattering Amplitudes in Gauge Theories}\href
  {http://arxiv.org/abs/1410.5256} {\path{arXiv:1410.5256}}.

\bibitem{Tye:2010dd}
S.~Henry~Tye, Y.~Zhang, {Dual Identities inside the Gluon and the Graviton
  Scattering Amplitudes}, JHEP 1006 (2010) 071.
\newblock \href {http://arxiv.org/abs/1003.1732} {\path{arXiv:1003.1732}},
  \href {http://dx.doi.org/10.1007/JHEP06(2010)071, 10.1007/JHEP04(2011)114}
  {\path{doi:10.1007/JHEP06(2010)071, 10.1007/JHEP04(2011)114}}.

\bibitem{Bern:1991aq}
Z.~Bern, D.~A. Kosower, {The Computation of loop amplitudes in gauge theories},
  Nucl. Phys. B379 (1992) 451--561.
\newblock \href {http://dx.doi.org/10.1016/0550-3213(92)90134-W}
  {\path{doi:10.1016/0550-3213(92)90134-W}}.

\bibitem{Bern:1995db}
Z.~Bern, A.~G. Morgan, {Massive loop amplitudes from unitarity}, Nucl. Phys.
  B467 (1996) 479--509.
\newblock \href {http://arxiv.org/abs/hep-ph/9511336}
  {\path{arXiv:hep-ph/9511336}}, \href
  {http://dx.doi.org/10.1016/0550-3213(96)00078-8}
  {\path{doi:10.1016/0550-3213(96)00078-8}}.

\bibitem{Bern:2002zk}
Z.~Bern, A.~De~Freitas, L.~J. Dixon, H.~L. Wong, {Supersymmetric
  regularization, two loop QCD amplitudes and coupling shifts}, Phys. Rev. D66
  (2002) 085002.
\newblock \href {http://arxiv.org/abs/hep-ph/0202271}
  {\path{arXiv:hep-ph/0202271}}, \href
  {http://dx.doi.org/10.1103/PhysRevD.66.085002}
  {\path{doi:10.1103/PhysRevD.66.085002}}.

\bibitem{Fazio:2014xea}
R.~A. Fazio, P.~Mastrolia, E.~Mirabella, W.~J. Torres~Bobadilla, {On the
  Four-Dimensional Formulation of Dimensionally Regulated Amplitudes}, Eur.
  Phys. J. C74~(12) (2014) 3197.
\newblock \href {http://arxiv.org/abs/1404.4783} {\path{arXiv:1404.4783}},
  \href {http://dx.doi.org/10.1140/epjc/s10052-014-3197-4}
  {\path{doi:10.1140/epjc/s10052-014-3197-4}}.

\bibitem{Mertig:1990an}
R.~Mertig, M.~Bohm, A.~Denner, {FEYN CALC: Computer algebraic calculation of
  Feynman amplitudes}, Comput. Phys. Commun. 64 (1991) 345--359.
\newblock \href {http://dx.doi.org/10.1016/0010-4655(91)90130-D}
  {\path{doi:10.1016/0010-4655(91)90130-D}}.

\bibitem{Maitre:2007jq}
D.~Maitre, P.~Mastrolia, {S@M, a Mathematica Implementation of the
  Spinor-Helicity Formalism}, Comput. Phys. Commun. 179 (2008) 501--574.
\newblock \href {http://arxiv.org/abs/0710.5559} {\path{arXiv:0710.5559}},
  \href {http://dx.doi.org/10.1016/j.cpc.2008.05.002}
  {\path{doi:10.1016/j.cpc.2008.05.002}}.

\bibitem{Zhu:1980sz}
D.-p. Zhu, {Zeros in Scattering Amplitudes and the Structure of Nonabelian
  Gauge Theories}, Phys.Rev. D22 (1980) 2266.
\newblock \href {http://dx.doi.org/10.1103/PhysRevD.22.2266}
  {\path{doi:10.1103/PhysRevD.22.2266}}.

\bibitem{DelDuca:1999rs}
V.~Del~Duca, L.~J. Dixon, F.~Maltoni, {New color decompositions for gauge
  amplitudes at tree and loop level}, Nucl. Phys. B571 (2000) 51--70.
\newblock \href {http://arxiv.org/abs/hep-ph/9910563}
  {\path{arXiv:hep-ph/9910563}}, \href
  {http://dx.doi.org/10.1016/S0550-3213(99)00809-3}
  {\path{doi:10.1016/S0550-3213(99)00809-3}}.

\bibitem{Bobadilla:2015wma}
W.~J.~T. Bobadilla, A.~R. Fazio, P.~Mastrolia, E.~Mirabella, {Generalised
  Unitarity for Dimensionally Regulated Amplitudes}\href
  {http://arxiv.org/abs/1505.05890} {\path{arXiv:1505.05890}}.

\bibitem{Jentschura:2012vp}
U.~D. Jentschura, B.~J. Wundt, {From Generalized Dirac Equations to a Candidate
  for Dark Energy}, ISRN High Energy Phys. 2013 (2013) 374612.
\newblock \href {http://arxiv.org/abs/1205.0521} {\path{arXiv:1205.0521}},
  \href {http://dx.doi.org/10.1155/2013/374612}
  {\path{doi:10.1155/2013/374612}}.

\bibitem{Hahn:2000kx}
T.~Hahn, {Generating Feynman diagrams and amplitudes with FeynArts 3}, Comput.
  Phys. Commun. 140 (2001) 418--431.
\newblock \href {http://arxiv.org/abs/hep-ph/0012260}
  {\path{arXiv:hep-ph/0012260}}, \href
  {http://dx.doi.org/10.1016/S0010-4655(01)00290-9}
  {\path{doi:10.1016/S0010-4655(01)00290-9}}.

\end{thebibliography}

\end{document}